# Quasi Normal Modes description of transmission properties for Photonic Band Gap structures.

*A. Settimi*[(1-2)], *S. Severini*[(3)], *B. J. Hoenders*[(4)]

[(1)] FILAS (Finanziaria Laziale di Sviluppo) – via A. Farnese 3, 00192 Roma, Italy
[(2)] INGV (Istituto Nazionale di Geofisica e Vulcanologia) –
via di Vigna Murata 605, 00143 Roma, Italy
[(3)] Centro Interforze Studi per le Applicazioni Militari –
via della Bigattiera 10, 56122 San Piero a Grado (Pi), Italy
[(4)] Research group theory of condensed matter
in the Institute for theoretical physics and Zernike Institute for advanced materials –
University of Groningen – Nijenborg 4, NL 9747 AG Groningen, Netherlands

**Abstract**

In this paper, we use the "Quasi Normal Modes" (QNM) approach for discussing the transmission properties of double-side opened optical cavities: in particular, this approach is specified for one dimensional (1D) "Photonic Band Gap" (PBG) structures. Moreover, we conjecture that the density of the modes (DOM) is a dynamical variable which has the flexibility of varying with respect to the boundary conditions as well as the initial conditions; in fact, the e.m. field generated by two monochromatic counter-propagating pump waves leads to interference effects inside a quarter-wave (QW) symmetric 1D-PBG structure. Finally, here, for the first time, a large number of theoretical assumptions on QNM metrics for an open cavity, never discussed in literature, are proved, and a simple and direct method to calculate the QNM norm for a 1D-PBG structure is reported.

# 1. Introduction.

Photonic crystals can be viewed as particular optical cavities having the properties of presenting allowed and forbidden bands for the electromagnetic radiation travelling inside, at optical frequencies. For these motivations these structures are also named "Photonic Band Gap" (PBG) [1]. In these structures, dispersive properties are usually evaluated assuming infinite periodic conditions [2]. The finite dimensions of PBG structures conceptually modify the calculation and the nature of the dispersive properties: this is mainly due to the existence of an energy flow into and out of the crystal. A phenomenological approach to the dispersive properties of one-dimensional (1D) PBG structures has been presented in ref. [3]. The application of the effective-medium approach is discussed, and the analogy with a simple Fabry-Perot structure is developed by Sipe et al. in ref. [4].

The problem of the field description inside an open cavity was discussed by several authors. In particular, Leung at al. [5] introduced the description of the electromagnetic field in a one side open microcavities in terms of "Quasi Normal Modes" (QNMs).
Microcavities are mesoscopic open systems. Since they are open, and therefore leaky, i.e. non-conservative, the resonance field eigen-functions are QNMs with complex eigen-frequencies. They play important roles in many optical processes. The analogy with normal modes of conservative systems is emphasized. In particular, in certain cases, the QNMs form a complete set, and much of the usual formalism can be carried through.
Microcavities are also open system, in which there has to be some output coupling. On account of the output coupling, the electromagnetic field and energy in the microcavity alone would be continuously lost to the outside. Thus, in physical terms, the microcavity is a non conservative system, while, from a mathematical standpoint, the operators which appear would be non-hermitian (or not self-adjoint). This then leads to interesting challenges in attempting to generalize the familiar tools of quantum mechanics and mathematical physics to such a non hermitian setting. The issues which arise, and the framework developed for addressing them, are generic to many other situations involving open systems.

Theoretical results were obtained in ref. [6] which developed a Hilbert-Schmidt type of theory, leading to a bilinear expansion in terms of the natural modes (QNMs) of the resolvent kernel connected with the integral equations of electromagnetic and potential scattering theory (see references therein [6]). The initial value problem considered by ref. [6] could be solved using this Hilbert-Schmidt type of theory taking the temporal Laplace transform of the Maxwell equations.
Applicative results were obtained by Bertolotti [7] who discussed the linear properties of one dimensional resonant cavities by using the matrix and the ray method (see references therein [7]).

Fabry-Perot, photonic crystals and 1D cavities in PBG structures are considered. QNMs for the description of the electromagnetic field in open cavities are introduced and some applications are given.

Theoretical and applicative results were obtained by ref. [8] which developed a scattering theory for finite 1D-PBG structures in terms of the natural modes (QNMs) of the scatterer. This theory generalizes the classical Hilbert-Schmidt type of bilinear expansions of the field propagator to a bilinear expansion into natural modes (QNMs) (see references therein [8]). It is shown that the Sturm-Liouville type of expansions for dispersive media differs considerably from those for non-dispersive media, they are e.g. overcomplete.

Conclusive theoretical and applicative results were obtained by Maksimovic [9] who used QNMs to characterize transmission resonances in 1D optical defect cavities and the related field approximations. Ref. [9] specializes to resonances inside the bandgap of the periodic multilayer mirrors which enclose the defect cavities. Using a field model with the most relevant QNMs, a variational principle permits to represent the field and the spectral transmission close to resonances.

In refs. [10], for the first time, the QNM approach was used and extended to the description of the scalar field behaviour in double-side opened optical cavities, in particular 1D-PBG structures. The validity of the approach is discussed by proving the QNM completeness, discussing the complex frequencies distribution, as well as the corresponding field distributions, and recovering the behaviour of the density of modes (DOM).

In ref. [11], the electromagnetic field inside an optical open cavity was analyzed in the framework of the QNM theory. The role of the complex frequencies in the transmission coefficient and their link with the DOM is clarified. An application to a quarter-wave (QW) symmetric 1D-PBG structure is discussed to illustrate the usefulness and the meaning of the results.

In ref. [12], by using the QNM formalism in a second quantization scheme, the problem of the counter-propagation of electromagnetic fields inside optical open cavities was studied. The links between QNM operators and canonical destruction and creation operators describing the external free field, as well as the field correlation functions, are found and discussed. An application of the theory is performed for open cavities whose refractive index satisfies symmetric properties.

In this paper, we use the QNM approach for discussing the transmission properties of double-side opened optical cavities: in particular, this approach is specified for 1D-PBG structures. Moreover, we conjecture that the DOM is a dynamical variable which has the flexibility of varying with respect to the boundary conditions as well as the initial conditions; in fact, the e.m. field generated by two monochromatic counter-propagating pump waves leads to interference effects inside a quarter-wave (QW) symmetric 1D-PBG structure. Finally, here, for the first time, a large number of theoretical assumptions on QNM metrics for an open cavity, never discussed in literature, are proved, and a simple and direct method to calculate the QNM norm for a 1D-PBG structure is reported.

This paper is organized as follows. In section 2, the QNM approach is introduced. In section 3, a large number of theoretical assumptions on QNM metrics for an open cavity, never discussed in literature, are proved. In section 4, the transmission coefficient for an open cavity is calculated as a superposition of QNMs. In section 5, the QNM approach is applied to a 1D-PBG structure and a transmission coefficient formula is obtained for a QW symmetric 1D-PBG structure. In section 6, the transmission resonances are compared with the QNM frequencies and the transmission "modes" at the resonances are calculated as super-positions of the QNM functions. In section 7, it has been shown that the e.m. field generated by two monochromatic counter-propagating pump waves leads to interference effects inside a QW symmetric 1D-PBG structure, such that a physical interpretation of the DOM is proposed. In section 8, a final discussion is reported and a comparison is proposed about all the obtained results and their associated theoretical improvements with respect to similar topics presented in literature (Maksimovic ref [9]). The Appendix describes a simple and direct method to calculate the QNM norm for a 1D-PBG structure.

## 2. Quasi Normal Mode (QNM) approach.

With reference to fig. 1, consider an open cavity as a region of length $L$, filled with a material of a given refractive index $n(x)$, which is enclosed in an infinite homogeneous external space. The cavity includes also the terminal surfaces, so it is represented as $C = [0, L]$ and the rest of universe as $U = (-\infty, 0) \cup (L, \infty)$.

The refractive index satisfies [5]:

- the *discontinuity conditions*, i.e. $n(x)$ presents a step at $x = 0$ and $x = L$, in this hypothesis a natural demarcation of a finite region is provided;
- the *no tail conditions*, i.e. $n(x) = n_0$ for $x < 0$ and $x > L$, in this hypothesis outgoing waves are not scattered back.

The e.m. field $E(x,t)$ in the open cavity satisfies the equation [13]

$$\left[\frac{\partial^2}{\partial x^2} - \rho(x)\frac{\partial^2}{\partial t^2}\right] E(x,t) = 0, \qquad (2.1)$$

where $\rho(x) = [n(x)/c]^2$, being $c$ the speed of light in vacuum. If there is no external pumping, the e.m. field satisfies suitable "outgoing waves" conditions [5][10]

$$\partial_x E(x,t) = \sqrt{\rho_0}\partial_t E(x,t) \quad \text{for} \quad x < 0, \tag{2.2}$$

$$\partial_x E(x,t) = -\sqrt{\rho_0}\partial_t E(x,t) \quad \text{for} \quad x > L, \tag{2.3}$$

where $\rho_0 = (n_0/c)^2$, being $n_0$ the outside refractive index. In fact, on the left side of the same cavity, i.e. $x < 0$, the e.m. field is travelling in the negative sense of the $x$-axis, i.e. $E(x,t) = E[x + (c/n_0)t]$, so eq. (2.2) holds as one can easily prove. On the right side of the cavity, i.e. $x > L$, the e.m. field is travelling in the positive sense of the $x$-axis, i.e. $E(x,t) = E[x - (c/n_0)t]$, so eq. (2.3) holds as well.

To take the cavity leakages into account, the Laplace transform of the e.m. field is considered [14],

$$\tilde{E}(x,\omega) = \int_0^\infty E(x,t)\exp(i\omega t)dt, \tag{2.4}$$

where $\omega$ is a complex frequency. The e.m. field has to satisfy the Sommerfeld radiative condition [13]:

$$\lim_{x \to \pm\infty} \tilde{E}(x,\omega) = 0. \tag{2.5}$$

Since $s = i\omega = \exp(i\pi/2)\cdot\omega$ with $\omega \in C$ [14], eq. (2.4) defines a transformation which looks like a Fourier transform with a complex frequency, but it is a $\pi/2$-rotated Laplace transform.

The $\pi/2$-rotated Laplace transform of the e. m. field converges to an analytical function $\tilde{E}(x,\omega)$ only over the half-plane of convergence $\text{Im}\,\omega > 0$. In fact, if the Laplace transform (2.4) is applied to the "outgoing waves conditions" (2.2), it follows

$$\partial_x \tilde{E}(x,\omega) = -i\omega\sqrt{\rho_0}\tilde{E}(x,\omega) \quad \text{for} \quad x < 0, \tag{2.6}$$

$$\partial_x \tilde{E}(x,\omega) = i\omega\sqrt{\rho_0}\tilde{E}(x,\omega) \quad \text{for} \quad x > L, \tag{2.7}$$

and, solving the last equation (2.7):

$$\tilde{E}(x,\omega) \propto \exp(i\omega\sqrt{\rho_0}x) = \exp(i\,\text{Re}\,\omega\sqrt{\rho_0}x)\exp(-\text{Im}\,\omega\sqrt{\rho_0}x) \quad \text{for} \quad x > L. \tag{2.8}$$

The Sommerfeld radiative condition (2.5) can be satisfied only if $\text{Im}\,\omega > 0$.

The transformed Green function $\tilde{G}(x,x',\omega)$ can be defined by [13]

$$\left[\frac{\partial^2}{\partial x^2} + \omega^2 \rho(x)\right]\tilde{G}(x,x',\omega) = -\delta(x-x'); \tag{2.9}$$

it is an e.m. field so, over the half-plane of convergence $\text{Im}\,\omega > 0$, it satisfies the Sommerfeld radiative conditions [13]:

$$\tilde{G}(x,x',\omega) \propto \begin{cases} \exp(i\omega\sqrt{\rho_0}x) \to 0 & \text{for} \quad x \to \infty \\ \exp(-i\omega\sqrt{\rho_0}x) \to 0 & \text{for} \quad x \to -\infty \end{cases}. \tag{2.10}$$

Two "auxiliary functions" $g_\pm(x,\omega)$ can be defined by [5]

$$\left[\frac{\partial^2}{\partial x^2}+\omega^2\rho(x)\right]g_\pm(x,\omega)=0 ; \qquad (2.11)$$

they are not defined as e.m. fields, because, over the half-plane of convergence $\text{Im}\,\omega>0$, they satisfy only the "asymptotic conditions" [5][10]:

$$\begin{cases} g_+(x,\omega)\propto \exp(i\omega\sqrt{\rho_0}x)\to 0 & \text{for } x\to\infty \\ g_-(x,\omega)\propto \exp(-i\omega\sqrt{\rho_0}x)\to 0 & \text{for } x\to-\infty \end{cases}. \qquad (2.12)$$

However, the transformed Green function $\tilde{G}(x,x',\omega)$ can be calculated in terms of the "auxiliary functions" $g_\pm(x,\omega)$. In fact, it can be shown that [5] the Wronskian $W(x,\omega)$ associated to the two "auxiliary functions" $g_\pm(x,\omega)$ is x-independent,

$$W(x,\omega)=g_+(x,\omega)g'_-(x,\omega)-g_-(x,\omega)g'_+(x,\omega)=W(\omega), \qquad (2.13)$$

and for the transformed Green function [5]:

$$\tilde{G}(x,x',\omega)=\begin{cases} -\dfrac{g_-(x,\omega)g_+(x',\omega)}{W(\omega)} & \text{for } x<x' \\ -\dfrac{g_+(x,\omega)g_-(x',\omega)}{W(\omega)} & \text{for } x'<x \end{cases}. \qquad (2.14)$$

In what follows it is proved that, just because of the "asymptotic conditions" (2.12), the "auxiliary functions" $g_\pm(x,\omega)$ are linearly independent over the half-plane of convergence $\text{Im}\,\omega>0$, and so the Laplace transformed Green function $\tilde{G}(x,x',\omega)$ is analytic over $\text{Im}\,\omega>0$.

The "asymptotic conditions" establish that, only over $\text{Im}\,\omega>0$, the auxiliary function $g_+(x,\omega)$ acts as an e.m. field for large $x$, because it is exponentially decaying. In fact, from eq. (2.12): $g_+(x,\omega)=\exp(i\omega\sqrt{\rho_0}x)=\exp(i\,\text{Re}\,\omega\sqrt{\rho_0}x)\exp(-\text{Im}\,\omega\sqrt{\rho_0}x)\to 0$ for $x\to\infty$. Then, still over $\text{Im}\,\omega>0$, the other auxiliary function $g_-(x,\omega)$ in general does not act as an e.m. field for large $x$, so it is exponentially increasing. In fact, according to eq. (2.12): $g_-(x,\omega)=A(\omega)\exp(i\omega\sqrt{\rho_0}x)+B(\omega)\exp(-i\omega\sqrt{\rho_0}x)$ for $x\to\infty$, with $B(\omega)\neq 0$, and so $g_-(x,\omega)\approx B(\omega)\exp(-i\omega\sqrt{\rho_0}x)=B(\omega)\exp(-i\,\text{Re}\,\omega\sqrt{\rho_0}x)\exp(\text{Im}\,\omega\sqrt{\rho_0}x)\to\infty$, for $x\to\infty$. It follows that the "auxiliary functions" $g_\pm(x,\omega)$ are linearly independent over $\text{Im}\,\omega>0$, because the Wronskian $W(\omega)$ is not null; in fact, from eq. (2.13): $W(\omega)=\lim_{x\to\infty}[g_+(x,\omega)g'_-(x,\omega)-g_-(x,\omega)g'_+(x,\omega)]=2i\sqrt{\rho_0}\omega B(\omega)\neq 0$ over $\text{Im}\,\omega>0$. Thus, the transformed Green function $\tilde{G}(x,x',\omega)$ is analytic over $\text{Im}\,\omega > 0$, where $\tilde{G}(x,x',\omega)$ does not diverge; in fact, from eq. (2.14): $\tilde{G}(x,x',\omega) \propto 1/W(\omega)$, with $W(\omega) \neq 0$ over $\text{Im}\,\omega > 0$.

For analytical continuation [14], the transformed Green function $\tilde{G}(x,x',\omega)$ can be extended also over the lower complex half-plane $\text{Im}\,\omega < 0$. According to ref. [15], it is always possible to define an infinite set of frequencies which are the poles of the transformed Green function $\tilde{G}(x,x',\omega)$, over the lower complex half-plane $\text{Im}\,\omega < 0$. In other words, there exists an infinite set of complex frequencies $\omega_n$, $n \in \mathbb{Z} = \{0, \pm 1, \pm 2, \ldots\}$, with negative imaginary parts $\text{Im}\,\omega_n < 0$, for which the Wronskian (2.13) is null [5]:

$$W(\omega_n) = 0. \tag{2.15}$$

The poles of the transformed Green function are referred to as Quasi-Normal-Mode (QNM) eigen-frequencies [15]. The definition of the QNM eigen-frequencies implies that the "auxiliary functions" $g_\pm(x,\omega)$ become linearly dependent when they are calculated at the QNM frequencies $\omega_n$, $n \in \mathbb{Z} = \{0, \pm 1, \pm 2, \ldots\}$; so, the auxiliary functions in the QNM's are such that [5]

$$g_+(x,\omega_n) = c(\omega_n) g_-(x,\omega_n) = f(x,\omega_n), \tag{2.16}$$

where $c(\omega_n)$ is a suitable complex constant. The above functions $f(x,\omega_n) = f_n(x)$ are referred as Quasi-Normal-Mode (QNM) eigen-functions [15]. The couples $[\omega_n, f_n(x)]$ are referred as *Quasi-Normal-Modes* because:

- they are characterized by complex frequencies $\omega_n$, so they are the *not-stationary modes* of an open cavity [5] (they are observed in the frequency domain as resonances of finite width or in the time domain as damped oscillations);
- under the "discontinuity and no tail conditions", the wave functions $f_n(x)$ form an *orthogonal basis only inside the open cavity* [5] (it is possible to describe the QNMs in a manner parallel to the normal modes of a closed cavity).

Applying the QNM condition (2.16) to the equation for the "auxiliary functions" (2.11), it follows that the QNMs $[\omega_n, f_n(x)]$ satisfy the equation [5]:

$$\left[\frac{d^2}{dx^2} + \omega_n^2 \rho(x)\right] f_n(x) = 0. \tag{2.17}$$

Moreover, applying the QNM condition (2.16) to the "asymptotic conditions" for the "auxiliary functions" (2.12), it follows that, only for long distances from the open cavity, the QNMs do not represent e.m. fields because they satisfy the QNM "asymptotic conditions" [5][10],

$$f_n(x) = \exp(\pm i\omega_n \sqrt{\rho_0} x) \to \infty \quad \text{for} \quad x \to \pm\infty, \tag{2.18}$$

while, inside the same cavity and near its terminal surfaces from outside, the QNMs represent not stationary modes, in fact the asymptotic conditions (2.18) imply the "formal" QNM "outgoing waves" conditions:

$$\partial_x f_n(x)\big|_{x=0} = -i\omega_n \sqrt{\rho_0} f_n(0), \tag{2.19}$$

$$\partial_x f_n(x)\big|_{x=L} = i\omega_n \sqrt{\rho_0} f_n(L). \tag{2.20}$$

The conditions (2.19)-(2.20) are called as "formal" because they are referred to the QNMs which do not represent e.m. fields for long distances from the open cavity, and as "outgoing waves" because they are formally identical to the real outgoing waves conditions for the e.m. field in proximity to the surfaces of the cavity. In fact, eqs. (2.19)-(2.20) for the QNMs can be derived if and only if the requirement of outgoing waves holds for the e.m. field.

An open system is not conservative because energy can escape to the outside. As a result, the time-evolution operator is not Hermitian in the usual sense and the eigenfunctions (factorized solutions in space and time) are no longer normal modes but quasi-normal modes (QNMs) whose frequencies ω are complex. QNM analysis has been a powerful tool for investigating open systems. Previous studies have been mostly system specific, and use a few QNMs to provide approximate descriptions.

In refs. [5], the authors review developments which lead to a unifying treatment. The formulation leads to a mathematical structure in close analogy to that in conservative, Hermitian systems. Hence many of the mathematical tools for the latter can be transcribed. Emphasis is placed on those cases in which the QNMs form a complete set and thus give an exact description of the dynamics.

More explicitly than refs. [5], we consider a Laplace transform of the e.m. field, to take the cavity leakages into account, and remark that, only in the complex domain defined by a $\pi/2$-rotated Laplace transform, the QNMs can be defined as the poles of the transformed Green function, with negative imaginary part. Our *iter* follows three steps:

1. the $\pi/2$-rotated Laplace transform of the e. m. field [see eqs. (2.4)-(2.5)] converges to an analytical function $\tilde{E}(x,\omega)$ only over the half-plane of convergence $\text{Im}\,\omega > 0$.

2. just because of the "asymptotic conditions" [see eq. (2.12)], the "auxiliary functions" $g_\pm(x,\omega)$ [see eq. (2.11)] are linearly independent over the half-plane of convergence $\text{Im}\,\omega > 0$, and so the Laplace transformed Green function $\tilde{G}(x,x',\omega)$ [see eq. (2.9)] is analytic over $\text{Im}\,\omega > 0$.

3. with respect refs. [5], we remark that: the transformed Green function $\tilde{G}(x,x',\omega)$ [see eq. (2.14)] can be extended also over the lower complex half-plane $\text{Im}\,\omega < 0$, for analytical continuation [14]; and, it is always possible to define an infinite set of frequencies which are the poles of the transformed Green function $\tilde{G}(x,x',\omega)$ [see eqs. (2.15)-(2.16)], over the lower complex half-plane $\text{Im}\,\omega < 0$, according to ref. [15].

## 3. QNM metrics.

The QNM norm is defined as [5]

$$\langle f_n | f_n \rangle = \frac{dW}{d\omega}\bigg|_{\omega=\omega_n}, \qquad (3.1)$$

and, with the method proposed for a one side open cavity [5], one can prove, for a both side open cavity, using the QNM condition (2.18) and the eqs. (2.11)-(2.12), that [5][10]:

$$\langle f_n | f_n \rangle = 2\omega_n \int_0^L \rho(x) f_n^2(x)\,dx + i\sqrt{\rho_0}\,[f_n^2(0) + f_n^2(L)]. \qquad (3.2)$$

Several remarks about this generalized norm are in order: it involves $f_n^2(x)$ rather then $|f_n(x)|^2$ and it is in general complex; it involves the two "terminal surface terms" $i\sqrt{\rho_0}\,f_n^2(0)$ and $i\sqrt{\rho_0}\,f_n^2(L)$. If the QNM function $f_n(x)$ is normalized, according to

$$f_n^N(x) = f_n(x)\sqrt{\frac{2\omega_n}{\langle f_n | f_n \rangle}}, \qquad (3.3)$$

then:

$$\langle f_n^N | f_n^N \rangle = 2\omega_n. \qquad (3.4)$$

The QNM inner product is defined as [5] [10]

$$\langle f_n^N | f_m^N \rangle = i\int_{0^+}^{L^-} \left[ f_n^N(x)\hat{f}_m^N(x) + \hat{f}_n^N(x)f_m^N(x) \right] dx + i\sqrt{\rho_0}\left[ f_n^N(0)f_m^N(0) + f_n^N(L)f_m^N(L) \right], \qquad (3.5)$$

if the QNM conjugate momentum $\hat{f}_n(x)$ is introduced, according to:

$$\hat{f}_n^N(x) = -i\omega_n \rho(x) f_n^N(x). \qquad (3.6)$$

One can prove that:
- the inner product (3.5) is in accordance with the QNM norm (3.2) [5];
- the QNMs form an orthogonal basis inside the open cavity [5], i.e. [10]

$$\langle f_n^N | f_m^N \rangle = \delta_{n,m}(\omega_n + \omega_m), \qquad (3.7)$$

being $\delta_{n,m}$ the Kronecker delta, i.e. $\delta_{n,m} = \begin{cases} 1, & n=m \\ 0, & n \neq m \end{cases}$.

Let us introduce the overlapping integral of the $n^{th}$ QNM

$$I_n = \int_0^L |f_n^N(x)|^2 \rho(x) dx, \qquad (3.8)$$

which is linked to the statistical weight in the density of modes (DOM) of the $n^{th}$ QNM [10]. If the open cavity is characterized by very slight leakages,

$$|\mathrm{Im}\,\omega_n| \ll |\mathrm{Re}\,\omega_n|, \qquad (3.9)$$

then the overlapping integral of the $n^{th}$ QNM converges to [10]:

$$I_n \cong 1. \qquad (3.10)$$

*Proof*:

$$I_n = \int_0^L |f_n^N(x)|^2 \rho(x) dx \stackrel{eq.(3.3)}{=}$$

$$= \left|\frac{2\omega_n}{\langle f_n | f_n \rangle}\right| \int_0^L |f_n(x)|^2 \rho(x) dx \stackrel{eq.(3.2)}{=} \qquad ; \qquad (3.11)$$

$$= \frac{\int_0^L |f_n(x)|^2 \rho(x) dx}{\left|\int_0^L f_n^2(x)\rho(x)dx + i\frac{\sqrt{\rho_0}}{2\omega_n}[f_n^2(0) + f_n^2(L)]\right|} \leq \frac{\int_0^L |f_n(x)|^2 \rho(x) dx}{\left|\int_0^L f_n^2(x)\rho(x)dx\right|} \leq 1$$

if $|\mathrm{Im}\,\omega_n| \ll |\mathrm{Re}\,\omega_n|$, then $f_n(x) \cong |f_n(x)|$, so eq. (3.10) holds.

Instead, if the open cavity is characterized by a some leakage, then the overlapping integral of the $n^{th}$ QNM can be calculated as [10]:

$$I_n = \frac{\sqrt{\rho_0}}{2|\mathrm{Im}\,\omega_n|}\left[|f_n^N(0)|^2 + |f_n^N(L)|^2\right]. \qquad (3.12)$$

*Proof*:

$$\langle f_n^N | f_m^N \rangle = i\int_0^L \left[f_n^N(x)\hat{f}_m^N(x) + \hat{f}_n^N(x)f_m^N(x)\right]dx + i\sqrt{\rho_0}\left[f_n^N(0)f_m^N(0) + f_n^N(L)f_m^N(L)\right] \stackrel{eq.(3.6)}{=}$$

$$= (\omega_n + \omega_m)\int_0^L f_n^N(x)\rho(x)f_m^N(x)dx + i\sqrt{\rho_0}\left[f_n^N(0)f_m^N(0) + f_n^N(L)f_m^N(L)\right] \stackrel{eq.(3.7)}{=}, \qquad (3.13)$$

$$= \delta_{n,m}(\omega_n + \omega_m)$$

so

$$\int_0^L f_n^N(x)\rho(x)f_m^N(x)dx = \delta_{n,m} - i\frac{\sqrt{\rho_0}}{\omega_n + \omega_m}[f_n^N(0)f_m^N(0) + f_n^N(L)f_m^N(L)]; \qquad (3.14)$$

if $m = -n$, then $\omega_{-n} = -\omega_n^*$, $f_{-n}(x) = f_n^*(x)$ and more $\delta_{n,-n} = 0$, so eq. (3.14) can be reduced to eqs. (3.8) and (3.12).

Finally, if the open cavity is characterized by very slight leakages [eq. (3.9)] and its refractive index $n(x)$ satisfies the symmetry property

$$n(L/2 - x) = n(L/2 + x), \qquad (3.15)$$

then the QNM norm can be approximated in modulus:

$$|\langle f_n | f_n \rangle| \cong 2\sqrt{\rho_0}\left|\frac{\omega_n}{\mathrm{Im}\,\omega_n}\right|. \qquad (3.16)$$

*Proof of eq. (3.16):*

$$I_n = \frac{\sqrt{\rho_0}}{2|\mathrm{Im}\,\omega_n|}\left[|f_n^N(0)|^2 + |f_n^N(L)|^2\right] \overset{eq.(3.3)}{=} \frac{\sqrt{\rho_0}}{|\langle f_n|f_n\rangle|}\left|\frac{\omega_n}{\mathrm{Im}\,\omega_n}\right|\left[|f_n(0)|^2 + |f_n(L)|^2\right] \overset{eq.(3.10)}{\cong} 1, \qquad (3.17)$$

so

$$|\langle f_n|f_n\rangle| \cong \sqrt{\rho_0}\left[|f_n(0)|^2 + |f_n(L)|^2\right]\left|\frac{\omega_n}{\mathrm{Im}\,\omega_n}\right| \overset{\mathrm{Hip:}\,f_n(0)=1}{=} \sqrt{\rho_0}\left[1 + |f_n(L)|^2\right]\left|\frac{\omega_n}{\mathrm{Im}\,\omega_n}\right|; \qquad (3.18)$$

if the open cavity is symmetric [eq. (3.15) holds], then $f_n(L) = (-1)^n f_n(0) = (-1)^n$, so eq. (3.18) can be reduced to eq. (3.16).

*Physical interpretation of eq. (3.16):*

- The modulus of QNM norm is expressed only in terms of the QNM frequencies;
- The modulus of the QNM norm is high ($|\langle f_n|f_n\rangle| \gg 2\sqrt{\rho_0}$) when the leakages of the open cavity are very slight ($|\mathrm{Im}\,\omega_n| \ll |\mathrm{Re}\,\omega_n|$);
- The QNM theory can be applied to open cavities and is based on the *outgoing waves* conditions [eqs. (2.2)-(2.3)] which formalize some leakages ($\mathrm{Im}\,\omega_n < 0$), so the QNM theory can not include the conservative case, when the cavities are closed and are not characterized by any leakages ($\mathrm{Im}\,\omega_n = 0$).

## 4. Calculation of transmission coefficient.

With reference to fig. 1.b., let us now consider an open cavity of length *L*, filled with a refractive index $n(x)$, in the presence of a pump incoming from the left side. The cavity includes also the surfaces, so it is represented as $C = [0, L]$ and the rest of universe as $U = (-\infty, 0) \cup (L, \infty)$. The refractive index satisfies the *discontinuity and no tail conditions* [5], as specified above.

Under these conditions, the QNMs form a complete basis only inside the cavity, and the e.m. field can be calculated as a superposition of QNMs [5]

$$E(x,t) = \sum_n a_n(t) f_n^N(x), \quad \text{for} \quad 0 \le x \le L, \tag{4.1}$$

where $f_n^N(x)$ are the normalized QNM functions [see eq. (3.3)].

The superposition coefficients $a_n(t)$ satisfy the dynamic equation [5]

$$\dot{a}_n(t) + i\omega_n a_n(t) = \frac{i}{2\omega_n \sqrt{\rho_0}} f_n^N(0) b(t), \tag{4.2}$$

where $b(t)$ is the driving force $b(t) = -2\sqrt{\rho_0}\, \partial_x E_P(x,t)\big|_{x=0}$.

The left-pump $E_P(x,t)$ satisfies the "incoming wave condition" [5]:

$$b(t) = -2\sqrt{\rho_0}\, \partial_x E_P(x,t)\big|_{x=0} = 2\rho_0 \partial_t E_P(0,t). \tag{4.3}$$

Each QNM is driven by the driving force $b(t)$ and at the same time decays because of $\operatorname{Im}\omega_n < 0$. The coupling to the force is determined by the surface value of the QNM wave function $f_n^N(0)$.

The cavity is in a steady state, so the Fourier transform with a real frequency $\tilde{E}(x,\omega) = \int_{-\infty}^{\infty} E(x,t) \exp(i\omega t) dt$ can be applied to equations (4.1)-(4.3), and it follows

$$\begin{cases} \tilde{E}(x,\omega) = \sum_n \tilde{a}_n(\omega) f_n^N(x) \\ \tilde{a}_n(\omega) = \dfrac{f_n^N(0)}{2\omega_n \sqrt{\rho_0}} \dfrac{\tilde{b}(\omega)}{\omega_n - \omega} \\ \tilde{b}(\omega) = -2i\rho_0 \omega \tilde{E}_P(0,\omega) \end{cases}, \tag{4.4}$$

so

$$\tilde{E}(x,\omega) = \tilde{E}_P(0,\omega) i\omega \sqrt{\rho_0} \sum_n \frac{f_n^N(0) f_n^N(x)}{\omega_n(\omega - \omega_n)}. \tag{4.5}$$

With reference to fig. 1.b., the e.m. field is continuous at cavity surfaces $x = 0$ and $x = L$, so $\tilde{E}(0^-,\omega) = \tilde{E}(0^+,\omega)$ and $\tilde{E}(L^-,\omega) = \tilde{E}(L^+,\omega)$, and the e.m. field $\tilde{E}(0,\omega)$ at surface $x = 0$ is the superposition of the incoming pump $\tilde{E}_P(0,\omega)$ and the reflected field $\tilde{E}_R(0,\omega)$, so $\tilde{E}(0,\omega) = \tilde{E}_P(0,\omega) + \tilde{E}_R(0,\omega)$, while the e.m. field $\tilde{E}(L,\omega)$ at the surface $x = L$ is only the transmitted field $\tilde{E}_T(L,\omega)$, so $\tilde{E}(L,\omega) = \tilde{E}_T(L,\omega)$.

It follows that the transmission coefficient $t(\omega)$ for an open cavity of length $L$ can be defined as the ratio between the transmitted field $\tilde{E}(L,\omega)$ at the surface $x = L$ and the incoming pump $\tilde{E}_P(0,\omega)$ at surface $x = 0$ [13]:

$$t(\omega) = \frac{\tilde{E}(L,\omega)}{\tilde{E}_P(0,\omega)}. \tag{4.6}$$

The transmission coefficient is obtained as superposition of QNMs inserting eq. (4.5) in eq. (4.6):

$$t(\omega) = i\omega\sqrt{\rho_0}\sum_n \frac{f_n^N(0)f_n^N(L)}{\omega_n(\omega-\omega_n)}. \tag{4.7}$$

Applying the QNM completeness condition [5] [19] $\sum_n \frac{f_n^N(x)f_n^N(x')}{\omega_n} = 0$ for $0 \leq x, x' \leq L$, the transmission coefficient simplifies as:

$$t(\omega) = i\sqrt{\rho_0}\sum_n \frac{f_n^N(0)f_n^N(L)}{\omega-\omega_n}. \tag{4.8}$$

Inserting the normalized QNM functions $f_n^N(x)$ [see eq.(3.3)], with $f_n(0) = 1$, equation (4.8) becomes:

$$t(\omega) = i2\sqrt{\rho_0}\sum_n \frac{f_n(L)}{\langle f_n | f_n \rangle}\frac{\omega_n}{\omega-\omega_n}. \tag{4.9}$$

For a symmetric cavity, such that $f_n(L) = (-1)^n f_n(0) = (-1)^n$, finally [11]

$$t(\omega) = \sum_n \frac{(-1)^n}{\gamma_n}\frac{\omega_n}{\omega-\omega_n}, \tag{4.10}$$

where $\gamma_n = \langle f_n | f_n \rangle / 2i\sqrt{\rho_0}$.

In accordance with ref. [11], the transmission coefficient $t(\omega)$ of an open cavity can be calculated as a superposition of suitable functions, with QNM norms $\langle f_n | f_n \rangle$ as weighting coefficients and QNM frequencies $\omega_n$ as parameters.

## 5. One dimensional Photonic Band Gap (1D-PBG) structures.

With reference to fig. 2, let us now consider a symmetric one dimensional (1D) Photonic Band Gap (PBG) structure which consists of $N$ periods plus one layer; every period is composed of two layers respectively with lengths $h$ and $l$ and with refractive indices $n_h$ and $n_l$, while the added layer is with parameters $h$ and $n_h$.

The symmetric 1D-PBG structure consists of $2N+1$ layers with a total length $L = N(h+l)+h$. If the two layers external to the symmetric 1D-PBG structure are considered, the 1D-space $x$ can be divided into $2N+3$ layers; they are $L_k = [x_k, x_{k+1}]$, $k = 0,1,\ldots,2N+1, 2N+2$, with $x_0 = -\infty$, $x_1 = 0$, $x_{2N+2} = L$, $x_{2N+3} = +\infty$.

The refractive index $n(x)$ takes a constant value $n_k$ in every layer $L_k$, $k = 0,1,\ldots,2N+1,2N+2$, i.e.

$$n(x) = \begin{cases} n_0 & \text{for } x \in L_0, L_{2N+2} \\ n_h & \text{for } x \in L_k, k = 1,3,\ldots,2N-1,2N+1 \\ n_l & \text{for } x \in L_k, k = 2,4,\ldots,2N \end{cases} \quad (5.1)$$

For the symmetric 1D-PBG structure with refractive index (5.1), the QNM norm $\langle f_n | f_n \rangle$ can be obtained in terms of the $\omega_n$ frequency and of the $A_{2N+2}(\omega)$, $B_{2N+2}(\omega)$ coefficients for the $g_-(x,\omega)$ auxiliary function in the $L_{2N+2}$ layer on the right side of the 1D-PBG structure (see Appendix):

$$\langle f_n | f_n \rangle = 2in_0 \frac{\omega_n}{c} A_{2N+2}(\omega_n) \left( \frac{dB_{2N+2}}{d\omega} \right)_{\omega=\omega_n}. \quad (5.2)$$

As proved in [10], for a quarter wave (QW) symmetric 1D-PBG structure with $N$ periods and $\omega_{ref}$ as reference frequency, there are $2N+1$ families of QNMs, i.e. $F_n^{QNM}$, $n \in [0,2N]$; the $F_n^{QNM}$ family of QNMs consists of infinite QNM frequencies, i.e. $\omega_{n,m}$, $m \in \mathbb{Z} = \{0,\pm 1,\pm 2,\ldots\}$, which have the same imaginary part, i.e. $\text{Im}\,\omega_{n,m} = \text{Im}\,\omega_{n,0}$, $m \in \mathbb{Z}$, and are lined with a step $\Delta = 2\omega_{ref}$, i.e. $\text{Re}\,\omega_{n,m} = \text{Re}\,\omega_{n,0} + m\Delta$, $m \in \mathbb{Z}$. It follows that, if the complex plane is divided into some ranges, i.e. $S_m = \{m\Delta < \text{Re}\,\omega < (m+1)\Delta\}$, $m \in \mathbb{Z}$, each of the QNM family $F_n^{QNM}$ drops only one QNM frequency over the range $S_m$, i.e. $\omega_{n,m} = (\text{Re}\,\omega_{n,0} + m\Delta,\ \text{Im}\,\omega_{n,0})$; there are $2N+1$ QNM frequencies over the range $S_m$ and they can be referred as $\omega_{n,m} = \omega_{n,0} + m\Delta$, $n \in [0,2N]$. Thus, there are $2N+1$ QNM frequencies over the basic range, i.e. $S_0 = \{0 < \text{Re}\,\omega < \Delta\}$; they correspond to $\omega_{n,0}$, $n \in [0,2N]$.

The QNM norm (5.2) becomes:

$$\langle f_{n,m} | f_{n,m} \rangle = 2in_0 \frac{\omega_{n,m}}{c} A_{2N+2}(\omega_{n,m}) \left( \frac{dB_{2N+2}}{d\omega} \right)_{\omega=\omega_{n,m}}. \quad (5.3)$$

The coefficients $A_{2N+2}(\omega)$, $B_{2N+2}(\omega)$ are of the type $\exp(i\delta)$ where $\delta = 2\pi(\omega/\omega_{ref})$ [10]. If coefficients $A_{2N+2}(\omega)$, $B_{2N+2}(\omega)$ are calculated at QNM frequencies $\omega_{n,m} = \omega_{n,0} + m\Delta$, where $\Delta = 2\omega_{ref}$, it follows $A_{2N+2}(\omega_{n,m}) = A_{2N+2}(\omega_{n,0})$ and $(dB_{2N+2}/d\omega)_{\omega=\omega_{n,m}} = (dB_{2N+2}/d\omega)_{\omega=\omega_{n,0}}$, so from eq. (5.3):

$$\gamma_{n,m} = \frac{\langle f_{n,m} | f_{n,m} \rangle}{2i(n_0/c)} = \frac{\gamma_{n,0}}{\omega_{n,0}}(\omega_{n,0} + m\Delta). \tag{5.4}$$

Thus, for the $F_n^{QNM}$ QNM family, the $\gamma_{n,m}$ norm is periodic with a step $\Gamma = \gamma_{n,0}(\Delta/\omega_{n,0})$.
The transmission coefficient (4.10) becomes

$$t(\omega) = \sum_{n=0}^{2N} \sum_{m=-\infty}^{\infty} \frac{(-1)^{n+m}}{\gamma_{n,m}} \frac{\omega_{n,0} + m\Delta}{\omega - \omega_{n,0} - m\Delta}, \tag{5.5}$$

and, taking into account eq. (5.4),

$$t(\omega) = \sum_{n=0}^{2N} (-1)^n \frac{\omega_{n,0}}{\gamma_{n,0}} \sum_{m=-\infty}^{\infty} (-1)^m \frac{1}{\omega - \omega_{n,0} - m\Delta}, \tag{5.6}$$

it converges to:

$$t(\omega) = \frac{\pi}{\Delta} \sum_{n=0}^{2N} (-1)^n \frac{\omega_{n,0}}{\gamma_{n,0}} \csc[\frac{\pi}{\Delta}(\omega - \omega_{n,0})]. \tag{5.7}$$

## 6. QNMs and transmission peaks of a symmetric 1D-PBG structure.

### 6.a. QNM frequencies and transmission resonances.

The transmission coefficient of a QW symmetric 1D-PBG structure, with $N$ periods and $\omega_{ref}$ as reference frequency, is calculated as a superposition of $2N+1$ QNM-cosecants, which correspond to the number of QNMs in $[0, 2\omega_{ref})$ range.

In fig. 3.a., the transmission coefficient predicted by QNM theory and then by numeric methods existing in the literature [16] are plotted for a QW symmetric 1D-PBG structure, where the reference wavelength is $\lambda_{ref} = 1\mu m$, the number of periods is $N = 6$ and the values of refractive indices are $n_h = 2$, $n_l = 1.5$. In figure 3.b., the two transmission coefficients are plotted for the same structure of fig. 3.a. but with $n_h = 3$, $n_l = 2$. In figure 3.c., the two transmission coefficients are plotted for the same structure as in figure 3.b., with an increased number of periods $N = 7$.
In accordance with ref. [11], we note the good agreement between the transmission coefficient predicted by the QNM theory and the one obtained by some numeric methods [16].

In the tables 1, the same examples of QW symmetric 1D-PBG structures are considered: (a) $\lambda_{ref} = 1\mu m$, $N = 6$, $n_h = 2$, $n_l = 1,5$, (b) $\lambda_{ref} = 1\mu m$, $N = 6$, $n_h = 3$, $n_l = 2$, (c) $\lambda_{ref} = 1\mu m$, $N = 7$, $n_h = 3$, $n_l = 2$. For each of the three examples, the low and high frequency band-edges are described in terms of their resonances ($\omega_{Band-Edge}/\omega_{ref}$) and phases $\angle t(\omega_{B.E.}/\omega_{ref})$; besides, there is one QNM close to every single band-edge: the real part of the QNM-frequency $\text{Re}(\omega_{QNM}/\omega_{ref})$ is reported together with the relative shift from the band-edge resonance $(\text{Re}\,\omega_{QNM} - \omega_{B.E.})/\omega_{B.E.}$.

From tables 1.a. and 1.b. ($N=6$), the low and high frequency band-edges are characterized by negative phases, which sum is nearly $(-\pi)$, while, from table 1.c. ($N=7$), their respective phases are positive with a sum $\pi$ approximately. In fact, a QW symmetric 1D-PBG structure, with $N$ periods and $\omega_{ref}$ as reference frequency, presents a transmission spectrum with $2N+1$ transmission resonances in the $[0, 2\omega_{ref})$ range, i.e. $\exists \omega_k^{peak} \Rightarrow t(\omega_k^{peak}) = \exp(i\varphi_k)$, $k = 0,1,\ldots,2N$. Only by numerical calculations (MATLAB) on eq. (5.7), one can succeed in proving that:

$$\begin{cases} \varphi_0 \cong 0 \\ \varphi_k + \varphi_{(2N+1)-k} \cong (-1)^{k+1}\pi, \ k=1,\ldots,N \end{cases} \tag{6.1}$$

So, the lowest resonance $\omega_0^{peak}$ in the range $[0, 2\omega_{ref})$ has always a phase $\varphi_0 \cong 0$. The phases $\varphi_N$ and $\varphi_{N+1}$ of the low and high frequency band-edges $\omega_N^{peak}$ and $\omega_{N+1}^{peak}$ are such that $\varphi_N + \varphi_{N+1} \cong (-1)^{N+1}\pi$; so, for a QW symmetric 1D-PBG structure with $N$ periods, the phases of the two band-edges are such that their sum is nearly $\pi$ when $N$ is an odd number or $(-\pi)$ approximately if $N$ is an even number.

Again from tables 1, there is one QNM close to every single band-edge, with a relative shift from the band-edge [11]. In fact, a QW symmetric 1D-PBG structure, with $N$ periods and $\omega_{ref}$ as reference frequency, presents a transmission coefficient which is a superposition of $2N+1$ QNM-cosecants (5.7), centred to the $2N+1$ QNM frequencies in $[0, 2\omega_{ref})$ range, i.e. $\omega_k^{QNM}$, $k=0,1,\ldots,2N$. The QNM-cosecants are not sharp functions, so, when they are superposed, an aliasing occurs; the peak of the $k^{th}$ QNM cosecant can feel the tails of the $(k-1)^{th}$ and $(k+1)^{th}$ QNM-cosecants, so the $k^{th}$ resonance of the transmission spectrum results effectively shifted from the $k^{th}$ QNM frequency i.e. $\Delta\omega_k = \text{Re}\,\omega_{k,0} - \omega_k^{peak} \neq 0, \ \forall k = 0,1,\ldots,2N$.

Comparing values in the tables 1.a. ($\Delta n = n_h - n_l = 0.5$) and table 1.b. ($\Delta n = 1$), the relative shift $(\text{Re}\,\omega_{QNM} - \omega_{B.E.})/\omega_{B.E.}$ decreases with the refractive index step $\Delta n$, while, comparing values in tables 1.b. ($N=6$) and 1.c. ($N=7$), the same shift increases with the number of periods $N$. In accordance with ref. [11], some numerical simulations prove that, the more the 1D-PBG structure presents a large number of periods ($L \gg \lambda_{ref}$) with an high refractive index step ($\Delta n \gg 0$), the more the $k^{th}$ QNM describes the $k^{th}$ transmission peak in the sense that $\mathrm{Re}\,\omega_{k,0}$ comes near to $\omega_k^{peak}$.

**6.b. QNM functions and transmission modes.**

Let us consider one monochromatic pump $\tilde{E}_p(x,\omega)$, which is incoming from the left side with unit amplitude and zero initial phase

$$\tilde{E}_p(x,\omega) \doteq \tilde{p}_\omega(x) = \exp(in_0 \frac{\omega}{c} x), \qquad (6.2)$$

being $n_0$ is the refractive index of the universe.

The e.m. field $\tilde{E}(x,\omega)$, inside the QW symmetric 1D-PBG structure, can be calculated as a superposition of QNMs [see sections 3 and 4]:

$$\tilde{E}(x,\omega) \doteq \tilde{E}_\omega(x) =$$
$$\overset{\text{QNMs}}{=} \tilde{p}_\omega(0) i\omega \sqrt{\rho_0} \sum_n \frac{f_n^N(0) f_n^N(x)}{\omega_n(\omega - \omega_n)} = i\sqrt{\rho_0} \sum_n \frac{f_n^N(0) f_n^N(x)}{\omega - \omega_n} = i2\sqrt{\rho_0} \sum_n \frac{f_n(x)}{\langle f_n | f_n \rangle} \frac{\omega_n}{\omega - \omega_n} = \sum_n \frac{f_n(x)}{\gamma_n} \frac{\omega_n}{\omega - \omega_n} =$$
$$\overset{\text{QW symmetric 1D-PBG structure}}{=} \sum_{n=0}^{2N} \sum_{m=-\infty}^{\infty} \frac{f_{n,m}(x)}{\gamma_{n,m}} \frac{\omega_{n,0} + m\Delta}{\omega - \omega_{n,0} - m\Delta} = \sum_{n=0}^{2N} \frac{\omega_{n,0}}{\gamma_{n,0}} \sum_{m=-\infty}^{\infty} \frac{f_{n,m}(x)}{\omega - \omega_{n,0} - m\Delta}$$
(6.3)

If the pump (6.2), incoming from the left side, is tuned at the transmission peak which is close to the $k^{th}$ QNM of the $[0, 2\omega_{ref})$ range, i.e. $\omega_k^{peak} \approx \mathrm{Re}\,\omega_{k,0}$ with $k = 0,1,\ldots 2N$, then the transmission "mode" $\tilde{E}_k(x)$ can be approximated as a superposition only of the dominant terms

$$\tilde{E}_k(x) = \sum_{n=0}^{2N} \sum_{m=-\infty}^{\infty} \tilde{a}_{n,m}(\omega_k^{peak}) f_{n,m}(x) =$$
$$= \sum_{m=-\infty}^{\infty} \tilde{a}_{k,m}(\omega_k^{peak}) f_{k,m}(x) + \sum_{n \neq k} \sum_{m=-\infty}^{\infty} \tilde{a}_{n,m}(\omega_k^{peak}) f_{n,m}(x) \cong, \qquad (6.4)$$
$$\cong \sum_{m=-\infty}^{\infty} \tilde{a}_{k,m}(\omega_k^{peak}) f_{k,m}(x)$$

where

$$\tilde{a}_{n,m}(\omega_k^{peak}) = \frac{\omega_{n,0} / \gamma_{n,0}}{\omega_k^{peak} - \omega_{n,0} - m\Delta}, \quad n = 0,1,\ldots 2N+1, \quad m \in \mathbb{Z} = \{0, \pm 1, \pm 2, \ldots\}. \qquad (6.5)$$

In fact, if the superposition (6.4)-(6.5) is calculated at the frequency $\omega_k^{peak} \approx \mathrm{Re}\,\omega_{k,0}$, then each dominant term is characterized by a coefficient $\tilde{a}_{k,m}(\omega_k^{peak})$ with a denominator including the addendum $\omega_k^{peak} - \omega_{k,0} = (\omega_k^{peak} - \text{Re}\,\omega_{k,0}) - i\,\text{Im}\,\omega_{k,0} \approx -i\,\text{Im}\,\omega_{k,0}$, so $\left|\tilde{a}_{k,m}(\omega_k^{peak})\right| \leq 1/\left|\omega_k^{peak} - \omega_{k,0}\right| \approx 1/\left|\text{Im}\,\omega_{k,0}\right| \to \infty$, being $k = 0,1,\ldots 2N+1$, while each neglected term is characterized by a coefficient $\tilde{a}_{n,m}(\omega_k^{peak})$ with a denominator including the addendum $\left|\omega_k^{peak} - \omega_{n,0}\right| >> \left|\omega_k^{peak} - \omega_{k,0}\right|$, so $\left|\tilde{a}_{n,m}(\omega_k^{peak})\right| < 1/\left|\omega_k^{peak} - \omega_{n,0}\right| << 1/\left|\omega_k^{peak} - \omega_{k,0}\right| \leq \left|\tilde{a}_{k,m}(\omega_k^{peak})\right|$, being $n \neq k$.

Thus, the e.m. "mode" $\tilde{E}_k(x)$ inside the QW symmetric 1D-PBG structure, tuned at the transmission peak $\omega_k^{peak}$ close to the $k^{th}$ QNM of the $[0, 2\omega_{ref})$ range, being $k = 0,1,\ldots 2N$, can be approximated as the superposition of the QNM functions $f_{k,m}(x)$ which belong to the $k^{th}$ QNM-family, being $m \in \mathbb{Z} = \{0, \pm 1, \pm 2, \ldots\}$; moreover, the weigh-coefficients of the superposition $\tilde{a}_{k,m}(\omega_k^{peak})$ are calculated in the transmission resonance $\omega_k^{peak}$ and depend from the $k^{th}$ QNM-family.

By numerical calculations (MATLAB) one on eqs. (6.4)-(6.5) one can prove that, when the transmission peak is closer to the QNM $\omega_k^{peak} \simeq \text{Re}\,\omega_{k,0}$, it is sufficient to calculate the transmission mode (6.4) by the I order approximation [9]

$$\tilde{E}_k(x) \simeq \tilde{a}_{k,0}(\omega_k^{peak}) f_{k,0}(x), \qquad (6.6)$$

where [9]

$$\tilde{a}_{k,0}(\omega_k^{peak}) = \frac{\omega_{k,0}/\gamma_{k,0}}{\omega_k^{peak} - \omega_{k,0}} \quad , \quad k = 0,1,\ldots 2N+1, \qquad (6.7)$$

otherwise, by the II order approximation, more refined,

$$\tilde{E}_k(x) \simeq \tilde{a}_{k,0}(\omega_k^{peak}) f_{k,0}(x) + \\ + \tilde{a}_{k,-1}(\omega_k^{peak}) f_{k,-1}(x) + \tilde{a}_{k,+1}(\omega_k^{peak}) f_{k,+1}(x) \qquad (6.8)$$

where

$$\tilde{a}_{k,\pm 1}(\omega_k^{peak}) = \frac{\omega_{k,0}/\gamma_{k,0}}{\omega_k^{peak} - (\omega_{k,0} \pm \Delta)} \quad , \quad k = 0,1,\ldots 2N+1. \qquad (6.9)$$

In accordance with ref. [9], in the hypothesis that $\omega_k^{peak} \simeq \text{Re}\,\omega_{k,0}$, it follows $\left|\omega_k^{peak} - \omega_{k,0}\right| << \left|\omega_k^{peak} - (\omega_{k,0} \pm \Delta)\right| < \ldots < \left|\omega_k^{peak} - (\omega_{k,0} + m\Delta)\right| < \ldots$, so $\left|\tilde{a}_{k,0}(\omega_k^{peak})\right| >> \left|\tilde{a}_{k,\pm 1}(\omega_k^{peak})\right| > \ldots > \left|\tilde{a}_{k,m}(\omega_k^{peak})\right| > \ldots$, being $|m| \geq 2$: in fact, the e.m. "mode" (6.4)-(6.5) tuned at the transmission peak $\omega_k^{peak}$, for the I order, can be approximated by eq. (6.6), corresponding to the $k^{th}$ QNM of the $[0, 2\omega_{ref})$ range (the resonance $\omega_k^{peak}$ is close enough to the QNM frequency $\text{Re}\,\omega_{k,0}$) (see ref. [9]); then, for the II order, by eq. (6.8), including the two adjacent $(k \pm 1)^{th}$ QNMs (if the resonance $\omega_k^{peak}$ is far from the QNM

frequencies $\text{Re}\,\omega_{k,\pm1} = \omega_{k,0} \pm \Delta$, but the $\tilde{a}_{k,0}(\omega)$ function tuned at $\omega_k^{peak}$ feels the tails of the $\tilde{a}_{k,\pm1}(\omega)$ functions tuned in $\text{Re}\,\omega_{k,\pm1}$); finally, the next orders can be neglected (because the resonance $\omega_k^{peak}$ is too far from the QNM frequencies $\text{Re}\,\omega_{k,m} = \omega_{k,0} + m\Delta$, so the $\tilde{a}_{k,0}(\omega)$ function tuned at $\omega_k^{peak}$ does not feel the tails of the $\tilde{a}_{k,m}(\omega)$ functions tuned in $\text{Re}\,\omega_{k,m}$, being $|m| \geq 2$).

By numerical calculations (MATLAB) on eqs. (6.6) and (6.8) one can prove that, if the QNM is even closer to the transmission peak of a QW symmetric 1D-PBG structure, i.e. $\omega_k^{peak} \cong \text{Re}\,\omega_{k,0}$, which is characterized by a large number of periods, i.e. $L \gg \lambda_{ref}$, and an high refractive index step, i.e. $\Delta n \gg 0$, then the I order approximation (6.6)-(6.7) can be reduced to the QNM approximation:

$$\tilde{E}_k(x) \approx f_{k,0}(x). \tag{6.10}$$

In fact, if $\omega_k^{peak} \cong \text{Re}\,\omega_{k,0}$, then eq. (6.7) can be approximated as

$$\tilde{a}_{k,0}(\omega_k^{peak}) = \frac{\omega_{k,0}/\gamma_{k,0}}{(\omega_k^{peak} - \text{Re}\,\omega_{k,0}) - i\,\text{Im}\,\omega_{k,0}} \approx \frac{\omega_{k,0}/\gamma_{k,0}}{-i\,\text{Im}\,\omega_{k,0}} \quad , \quad k = 0,1,\ldots 2N+1; \tag{6.11}$$

if $L \gg \lambda_{ref}$, $\Delta n \gg 0$, then the 1D-PBG structure is characterized by very slight leakages [see eq. (3.9)], and, since the refractive index satisfies suitable symmetry properties [see eq. (3.15)], the approximated QNM norm (3.16) can be used

$$|\gamma_{k,0}| = \frac{|\langle f_{k,0} | f_{k,0} \rangle|}{2\sqrt{\rho_0}} \cong \left|\frac{\omega_{k,0}}{\text{Im}\,\omega_{k,0}}\right| \quad , \quad k = 0,1,\ldots 2N+1 \tag{6.12}$$

and, in modulus, the weigh-coefficient (6.11) converges to:

$$|\tilde{a}_{k,0}(\omega_k^{peak})| \approx \frac{1}{|\gamma_{k,0}|} \left|\frac{\omega_{k,0}}{\text{Im}\,\omega_{k,0}}\right| \cong 1 \quad , \quad k = 0,1,\ldots 2N+1. \tag{6.13}$$

Thus, in accordance with ref. [11], some numerical simulations prove that the more the 1D-PBG structure presents a large number of periods with an high refractive index step, the more the QNMs describe the transmission peaks in the sense that the QNM functions approximate the e.m. modes in the transmission resonances.

In fig. 4, with reference to a QW symmetric 1D-PBG structure ($\lambda_{ref} = 1\mu m$, $N = 6$, $n_h = 3$, $n_l = 2$), the e.m. "mode" intensities (a) $I_{I\,band-edge}$ at the low frequency band edge ($\omega_{I\,band-edge}/\omega_{ref} = 0.822$) and (b) $I_{II\,band-edge}$ at the high frequency band edge ($\omega_{II\,band-edge}/\omega_{ref} = 1.178$), in units of the intensity for an incoming pump $I_{pump}$, are plotted as functions of the dimensionless space $x/L$, where $L$ is the length of the 1D-PBG structure. It is clear the shifting between the QNM approximation (6.10) (the e.m. mode intensity $I_{I\,band-edge}$ is due to QNM close to the band-edge) and the first order approximation (6.6)-(6.7) [more refined than (6.10), $I_{band-edge}$ is calculated as the product of the QNM, close to the frequency band-edge, for a weigh coefficient, which takes into account the shift between the band-edge resonance and the QNM frequency] or the second order approximation (6.8)-(6.9) [even more refined, but almost superimposed to (6.6)-(6.7), $I_{band-edge}$ is calculated as the sum of the first order contribution due to the QNM, close to the frequency band-edge, and the second order contributions of the two adjacent QNMs, with the same imaginary part, so belonging to the same QNM family].

## 7. Quarter wave (QW) symmetric 1D-PBG structures excited by two counter-propagating field pumps.

Let us consider a QW symmetric 1D-PBG structure with $N$ periods and $\omega_{ref}$ as reference frequency. If the universe includes the terminal surfaces of the 1D-PBG, the cavity is represented as $C = [0, L]$ and the rest of universe as $U = (-\infty, 0) \cup (L, \infty)$. The 1D-PBG presents a refractive index $n(x)$ which satisfies the symmetry properties $n(L/2 - x) = n(L/2 + x)$. The cavity is characterized by a transmission spectrum with $2N + 1$ resonances in the $[0, 2\omega_{ref})$ range [10], i.e. $\exists\, \omega_k \Rightarrow t(\omega_k) = \exp[i\varphi_k],\ r^{\rightleftarrows}(\omega_k) = 0,\ k = 0, 1, \ldots 2N$ [17].

Let us consider one pump, coming from the left side, which is tuned at the $k^{th}$ resonance of the 1D-PBG structure, with unit amplitude and zero initial phase

$$\tilde{p}_k^{(\rightarrow)}(x) = \exp(i n_0 \frac{\omega_k}{c} x),\qquad(7.1)$$

being $n_0$ the refractive index of the universe. It excites in the cavity an e.m. "mode" $\tilde{E}_k^{(\rightarrow)}(x)$ which satisfies the equation [13]

$$\frac{d^2 \tilde{E}_k^{(\rightarrow)}}{dx^2} + \left(\frac{\omega_k}{c}\right)^2 n^2(x) \tilde{E}_k^{(\rightarrow)}(x) = 0,\qquad(7.2)$$

and the boundary conditions are [13]:

$$\begin{cases} \tilde{E}_k^{(\rightarrow)}(0) = \tilde{p}_k^{(\rightarrow)}(0) + r^{(\rightarrow)}(\omega_k) \tilde{p}_k^{(\rightarrow)}(0) = \tilde{p}_k^{(\rightarrow)}(0) = 1 \\ \tilde{E}_k^{(\rightarrow)}(L) = t(\omega_k) \tilde{p}_k^{(\rightarrow)}(0) = t(\omega_k) \end{cases}.\qquad(7.3)$$

Let us consider another pump, coming from the right side, tuned at the $k^{th}$ resonance of the 1D-PBG structure with unit amplitude and constant phase-difference $\Delta\varphi$, i.e.

$$\tilde{p}_k^{(\leftarrow)}(x) = \exp[-in_0(\omega_k/c)(x-L)]\exp[-i\Delta\varphi]. \tag{7.4}$$

If alone, this pump excites in the cavity an e.m. "mode" $\tilde{E}_k^{(\leftarrow)}(x)$ which satisfies an equation similar to (7.2) [13]

$$\frac{d^2\tilde{E}_k^{(\leftarrow)}}{dx^2} + \left(\frac{\omega_k}{c}\right)^2 n^2(x)\tilde{E}_k^{(\leftarrow)}(x) = 0, \tag{7.5}$$

because of the symmetry properties of the refractive index, with boundary conditions which are different from (7.3) [13]:

$$\begin{cases} \tilde{E}_k^{(\leftarrow)}(L) = \tilde{p}_k^{(\leftarrow)}(L) + r^{(\leftarrow)}(\omega_k)\tilde{p}_k^{(\leftarrow)}(L) = \tilde{p}_k^{(\leftarrow)}(L) = \exp[-i\Delta\varphi] \\ \tilde{E}_k^{(\leftarrow)}(0) = t(\omega_k)\tilde{p}_k^{(\leftarrow)}(L) = t(\omega_k)\exp[-i\Delta\varphi] \end{cases}. \tag{7.6}$$

It is easy to verify that the e.m. mode $\tilde{E}_k^{(\leftarrow)}(x)$, which is excited by the pump (7.4) and solves the equation (7.5) with the boundary conditions (7.6), is related to the e.m. mode $\tilde{E}_k^{(\rightarrow)}(x)$, which is excited by the pump (7.1) and solves equation (7.2) with the boundary conditions (7.3), by the link:

$$\tilde{E}_k^{(\leftarrow)}(x) = [\tilde{E}_k^{(\rightarrow)}(x)]^* t(\omega_k)\exp[-i\Delta\varphi]. \tag{7.7}$$

The two counter-propagating pumps, which are tuned at the $k^{th}$ resonance of the 1D-PBG structure, excite an e.m. "mode" $\tilde{E}_k(x)$ inside the cavity, which is the linear super-position of $\tilde{E}_k^{(\rightarrow)}(x)$ and $\tilde{E}_k^{(\leftarrow)}(x)$, i.e.[13]

$$\tilde{E}_k(x) = \tilde{E}_k^{(\rightarrow)}(x) + \tilde{E}_k^{(\leftarrow)}(x) = \tilde{E}_k^{(\rightarrow)}(x) + [\tilde{E}_k^{(\rightarrow)}(x)]^* t(\omega_k)\exp[-i\Delta\varphi]. \tag{7.8}$$

It is easy to calculate from (7.8) the e.m. mode intensity inside the cavity [13]

$$\begin{aligned} \tilde{I}_k(x) &= \tilde{E}_k(x)[\tilde{E}_k(x)]^* = \\ &= 2\tilde{I}_k^{(\rightarrow)}(x) + 2\operatorname{Re}\{[\tilde{E}_k(x)]^2 t^*(\omega_k)\exp[i\Delta\varphi]\} = \\ &= 2\tilde{I}_k^{(\rightarrow)}(x)\{1+\cos[2\phi_k^{(\rightarrow)}(x)+\Delta\varphi-\varphi_k]\} = \\ &= 4\tilde{I}_k^{(\rightarrow)}(x)\cos^2[\phi_k^{(\rightarrow)}(x)+\frac{\Delta\varphi-\varphi_k}{2}] \end{aligned}, \tag{7.9}$$

if the e.m. mode excited by the pump (7.1) is represented by $\tilde{E}_k^{(\rightarrow)}(x) = \sqrt{\tilde{I}_k^{(\rightarrow)}(x)}\exp[i\phi_k^{(\rightarrow)}(x)]$.
Eq. (7.9) shows that the e.m. field generated by two monochromatic counter-propagating pump waves leads to interference effects inside a QW symmetric 1D-PBG structure.

### 7.1. Density of modes (DOM).

In fig. 5, with reference to a symmetric QW 1D-PBG structure ($\lambda_{ref} = 1\mu m$, $N = 6$, $n_h = 3$, $n_l = 2$), the "e.m. mode" intensity excited inside the open cavity by one field pump (- - -) coming from the left side [I order approximation (6.6)] and by two counter-propagating field pumps in phase (—) [see eq. (7.9)] are compared when each of the two field pumps are tuned at (a) the low frequency band-edge ($\omega_{I\,band-edge}/\omega_{ref} = 0.822$) or (b) the high frequency band-edge ($\omega_{II\,band-edge}/\omega_{ref} = 1.178$). The e.m. mode intensities $I_{I\,band-edge}$ and $I_{II\,band-edge}$, in units of the intensity for the two pumps $I_{pump}$, are plotted as functions of the dimensionless space $x/L$, where $L$ is the length of the 1D-PBG structure. If the two counter-propagating field pumps in phase are tuned at the low frequency band-edge, there is a constructive interference and the field distribution in that band-edge is reminiscent of the distribution excited by one field pump in the same transmission peak; while, if the two pumps are tuned at the high frequency band-edge, there is a destructive interference and almost no e.m. field penetration occurs in the structure.

Some numerical simulations on eq. (7.9) prove that, when a QW symmetric 1D-PBG structure with $N$ periods is excited by two counter-propagating field pumps with a phase-difference $\Delta\varphi$, if $N$ is an even number, the e.m. mode intensities in the low (high) frequency band-edge increases in strength when the two field pumps are in phase $\Delta\varphi = 0$ (out of phase $\Delta\varphi = \pi$) and almost flags when the two pumps are out phase $\Delta\varphi = \pi$ (in phase $\Delta\varphi = 0$). Moreover, if $N$ is an odd number, the two field distributions in the low and high frequency band-edges exchange their physical response with respect to the phase-difference of the two pumps. The duality between the field distributions in the two band-edges, for which the one increases in strength when the other almost flags, can be explained recalling that the transmission phases of the two band edges are such that their sum is $\pi$ ($-\pi$) when $N$ is an odd (even) number.

Therefore, if a 1D-PBG structure is excited by two counter-propagating field pumps which are tuned at one transmission resonance, no e.m. "mode" can be produced because it depends not only on the boundary conditions (the two field pumps) but also on the initial conditions (the phase-difference of the two pumps) [12]. We conjecture that the "density of modes" is a dynamic variable which has the flexibility of varying with respect to the boundary conditions as well as the initial conditions, in accordance with ref. [12].

## 8. A final discussion.

The Quasi Normal Mode (QNM) theory considers the realistic situation in which an optical cavity is open from both sides and is enclosed in an infinite homogeneous external space; the lack of energy conservation for optical open cavities gives resonance field eigen-functions with complex eigen-frequencies. The time-space evolution operator for the cavity is not hermitian: the modes of the field are quasi-normal, i.e. they form an orthogonal basis only inside the cavity.

The behavior of the electromagnetic field in the optical domain, inside one-dimensional photonic crystals, is analyzed by using an extension of the QNM theory. One-dimensional (1D) Photonic Band Gap (PBG) structures are particular optical cavities, with both sides open to the external environment, with a stratified material inside. These 1D-PBG structures are finite in space and, when we work with electromagnetic pulses of a spatial extension longer than the length of the cavity, the 1D-PBG cannot be studied as an infinite structure: rather we have to consider the boundary conditions at the two ends of the cavity.

In the present paper, the QNM approach has been applied to a double-side opened optical cavity, and specified for a 1D-PBG structure. All the considerations made are exhaustively demonstrated and all the subsequently results agree with the ones presented in literature (see refs. [10] and [11]).

In this paper (sec. 2), suitable "outgoing wave" conditions for an open cavity without external pumping are formalized [see eqs. (2.2)-(2.3)]. More explicitly than refs. [5], here we have considered a Laplace transform of the e.m. field, to take the cavity leakages into account, and we have remarked that, only in the complex domain defined by a $\pi/2$-rotated Laplace transform, the QNMs can be defined as the poles of the transformed Green function, with negative imaginary part. Our *iter* has followed three steps:

1. the $\pi/2$-rotated Laplace transform of the e. m. field [see eqs. (2.4)-(2.5)] converges to an analytical function $\tilde{E}(x,\omega)$ only over the half-plane of convergence $\text{Im}\,\omega > 0$.

2. just because of the "asymptotic conditions" [see eq. (2.12)], the "auxiliary functions" $g_\pm(x,\omega)$ [see eq. (2.11)] are linearly independent over the half-plane of convergence $\text{Im}\,\omega > 0$, and so the Laplace transformed Green function $\tilde{G}(x,x',\omega)$ [see eq. (2.9)] is analytic over $\text{Im}\,\omega > 0$.

3. Differently from refs. [5], here we have remarked that: the transformed Green function $\tilde{G}(x,x',\omega)$ [see eq. (2.14)] can be extended also over the lower complex half-plane $\text{Im}\,\omega < 0$, for analytical continuation [14]; and, it is always possible to define an infinite set of frequencies which are the poles of the transformed Green function $\tilde{G}(x,x',\omega)$ [see eqs. (2.15)-(2.16)], over the lower complex half-plane $\text{Im}\,\omega < 0$, according to ref. [15].

Moreover (sec. 3), if the open cavity is characterized by very slight leakages and its refractive index satisfies suitable symmetry properties, more than refs. [5], here we have proved that the modulus of the QNM norm in modulus can be expressed only in terms of the QNM frequencies [see eq. (3.16)] and, more than refs. [10] and [11], here we have proposed its physical interpretation: the QNM norm is high when the leakages of the open cavity are very slight; the QNM theory can be applied to open cavities and is based on the "outgoing waves" conditions which formalize some leakages ($\text{Im}\,\omega_n < 0$), so the QNM theory can not include the conservative case, when the cavities are closed and are not characterized by any leakages ($\text{Im}\,\omega_n = 0$).

The paper (sec. 4) has given a proof that the transmission coefficient $t(\omega)$ can be calculated as a superposition of suitable functions, with QNM norms $\langle f_n | f_n \rangle$ as weighting coefficients and QNM frequencies $\omega_n$ as parameters [see eq.(4.10)], thesis only postulated in ref. [11].

The QNM approach has been applied to a 1D-PBG structure (sec. 5): more than ref. [11], here the transmission coefficient of a quarter wave (QW) symmetric 1D-PBG structure, with $N$ periods and $\omega_{ref}$ as reference frequency, has been calculated as a superposition of $2N+1$ QNM-cosecants, which correspond to the number of QNMs in $[0, 2\omega_{ref})$ range [see eq. (5.7)]; differently from the assertion of ref. [8] for which the frequency dependence of $t(\omega)$ could suggest at first glance that the positions of maxima of $T(\omega) = |t(\omega)|^2$ are realized when the real frequency coincides with the real part of the QNM frequencies $\omega_n$, here we have deduced that this fact is true only in the further hypothesis of real parts of QNM frequencies $\omega_n$ well separated on the real frequency axis, whereas, on the contrary, overlapping of the tails of two near QNM-cosecant contributions to $T(\omega)$ lead to a significantly displayed position of the maxima of $T(\omega)$, a sort of aliasing process (see figs. 4.a-c. and tab. 1.a-c.).

In sec. 6., the transmission resonances have been compared with the QNM frequencies (sec. 6.a.) and the transmission "modes" in the resonances have been calculated as super-positions of the QNM functions (sec. 6.b.): differently from Maksimovic paper [9], here we affirm explicitly that a proper field representation for the transmittance problem can be established in terms of QNMs, and the boundary conditions are not violated [see eqs. (6.2)-(6.3)]; even if the fields are entirely different when considered on the whole real axis, the e.m. field can be represented as a superposition of QNMs inside the open cavity and on the limiting surfaces too.

In the present paper (sec. 7), it has been shown that the e.m. field generated by two monochromatic counter-propagating pump waves leads to interference effects inside a QW symmetric 1D-PBG structure [see eq. (7.9)].

Even if this paper is chronologically subsequent to ref. [11], anyway it is conceived as a *trait d'union* between refs. [11] and [12], in fact the paper forestalls conceptually, according to classic electrodynamics, what ref. [12] develops in terms of quantum mechanics.

We have shown that if a 1D-PBG structure is excited by two counter-propagating pumps which are tuned at one transmission resonance, no e.m. mode can be produced because it depends not only on the boundary conditions (the two pumps) but also on the initial conditions (the phase-difference of the two pumps) [see fig. 5.a-b.]; in accordance with ref. [12], we have conjectured (sec 7.1) that the density of the modes (DOM) is a dynamical variable which has the flexibility to adjust with respect to the boundary conditions as well the initial conditions.

In Appendix, a direct method has been described to obtain the QNM norm for a QW symmetric 1D-PBG structure: the QNM norm $\langle f_n | f_n \rangle$ can be obtained in terms of the $\omega_n$ frequency and of the $A_{2N+2}(\omega)$, $B_{2N+2}(\omega)$ coefficients for the $g_-(x,\omega)$ auxiliary function in the $L_{2N+2}$ layer on the right side of the 1D-PBG structure.

**8.1. A comparison with Maksimovic ref [9].**

Let us propose a critical comparison between Maksimovic and our methods:
- Maksimovic specializes ref. [9] to finite periodic structures which possess transmission properties with a bandgap, i.e. with a region of frequencies of very low transmission. He chooses a field model for the transmittance problem and takes the relevant QNMs as those with the real part of their complex frequency in the given frequency range. To determine the decomposition coefficients in the field model, a variational form of the transmittance problem is employed. The transmittance problem corresponds to the equation and natural boundary conditions, arising form the condition of stationarity of a functional $F[E(x,t)]$. If $F[E(x,t)]$ becomes stationary, i.e. if the first variation of $F[E(x,t)]$ vanishes for arbitrary variations of the e.m. field $E(x,t)$, then $E(x,t)$ satisfies the natural boundary conditions. The optimal coefficients can be obtained as solutions of a system of linear equations. By solving the system for each value of the real frequency, the decomposition coefficients in the field representation for transmittance problem are obtained. This enables approximation of the spectral transmittance and reflectance and the related field profile.

  Maksimovic applies a method which is based on a mathematical principle, involving arduous calculations. Instead, we have proposed, for an open cavity, a physical method which starts from the boundary conditions at the surfaces of the cavity (sec. 4). Moreover, we have provided, for a symmetric 1D-PBG structure (sec. 5), a new algorithm to calculate the QNM norm by a derivative [see eq. (5.2)], more direct than the integral of ref. [9]; and, we have proved that a QW symmetric 1D-PBG structure with $N$ periods is characterized by $2N+1$ families of QNM frequencies, periodically distributed on the complex plane, such that, for each family, the QNM norm is periodical with a proper step [see eq.(5.4)]: as a result, the calculation of the transmission coefficient is even more simplified with respect to ref. [9].

- A common assumption made in the literature [5][7][10][18] is that the spectral transmission for single resonance situation, as described, is of a Lorentian like shape. We have fallen into line of this assumption (sec. 6.b): in fact, the e.m. field, inside a QW symmetric 1D-PBG structure, has been calculated as a superposition of QNMs; so, each single resonance is described by a Lorentian like shape [see eqs. (6.3), (6.4)-(6.5) and comments below].

  The method of ref. [9] justifies analytically this common assumption. Maksimovic considers the contribution of a single QNM in the field model. Here (sec. 6.b.), we have considered the contributions of the adjacent QNMs, by more and more refined approximations including successive orders: in fact, some numerical simulations have proved that (see figs. 4.a-b.), when the transmission peak is close to the QNM, it is sufficient to calculate the transmission mode by the I order approximation [see eqs. (6.6)-(6.7)], otherwise, by the II order approximation, more refined [see eqs. (6.8)-(6.9) and comments below]; so, in our opinion, the contribution of a single QNM is sufficient for the field model only in the limit of narrow resonances [see eqs. (6.10), (6.11)-(6.13) and comments below] .

## Appendix: QNM norm for 1D-PBG structures.

With reference to fig.2, let us now consider a symmetric one dimensional photonic band gap (1D-PBG) structure which consists of $N$ period plus one layer; every period is composed of two layers respectively with lengths $h$ and $l$ and with refractive indices $n_h$ and $n_l$, while the added layer is with parameters $h$ and $n_h$.

The symmetric 1D-PBG structure consists of $2N+1$ layers with a total length $L = N(h+l) + h$. If the two layers external to the symmetric 1D-PBG structure are considered, the 1D-space $x$ can be divided into $2N+3$ layers; the two external layers, $L_0 = (-\infty, 0)$ and $L_{2N+2} = (L, +\infty)$, and the 1D-PBG layers, the odd ones $L_{2k-1} = [(k-1)(h+l), (k-1)(h+l)+h]$, $k = 1, 2, \ldots, N+1$ and the even ones $L_{2k} = [(k-1)(h+l)+h, k(h+l)]$, $k = 1, 2, \ldots, N$.

The refractive index $n(x)$ takes a constant value $n_k$ in every layer $L_k$, $k = 0,1,\ldots,2N+1, 2N+2$, i.e.

$$n(x) = \begin{cases} n_0 & \text{for } x \in L_0, L_{2N+2} \\ n_h & \text{for } x \in L_k, k = 1,3,\ldots,2N-1, 2N+1 \\ n_l & \text{for } x \in L_k, k = 2,4,\ldots,2N \end{cases} \quad (A.1)$$

At first, the calculation of quasi normal modes (QNMs) is stated for the 1D-PBG structure with refractive index (A.1). The homogeneous equation is solved for the auxiliary functions (2.11), with the "asymptotic conditions" (2.12). The auxiliary function $g_-(x,\omega)$ is

$$\begin{aligned}
g_-(x,\omega) = & \left( A_0 e^{in_0 \frac{\omega}{c} x} + B_0 e^{-in_0 \frac{\omega}{c} x} \right) \vartheta(-x) + \\
& + \sum_{k=1}^{N+1} \left( A_{2k-1} e^{in_h \frac{\omega}{c} x} + B_{2k-1} e^{-in_h \frac{\omega}{c} x} \right) \vartheta[x-(k-1)(h+l)]\vartheta[(k-1)(h+l)+h-x] + \\
& + \sum_{k=1}^{N} \left( A_{2k} e^{in_l \frac{\omega}{c} x} + B_{2k} e^{-in_l \frac{\omega}{c} x} \right) \vartheta[x-(k-1)(h+l)-h]\vartheta[k(h+l)-x] + \\
& + \left( A_{2N+2} e^{in_0 \frac{\omega}{c} x} + B_{2N+2} e^{-in_0 \frac{\omega}{c} x} \right) \vartheta(x-L)
\end{aligned} \quad (A.2)$$

where $\vartheta(x)$ is the unit step function; while the auxiliary function $g_+(x,\omega)$ has a similar expression to (A.2), where some coefficients $C_k(\omega)$ and $D_k(\omega)$ replace $A_k(\omega)$ and $B_k(\omega)$ with $k=0,1,\ldots,2N+1,2N+2$. The auxiliary function $g_-(x,\omega)$ is a "right to left wave" for $x<0$, so $A_0(\omega)=0$, while the auxiliary function $g_+(x,\omega)$ is a "left to right wave" for $x>L$, so $D_{2N+2}(\omega)=0$.

If the continuity conditions are imposed at the 1D-PBG surfaces for the auxiliary function $g_-(x,\omega)$ and its spatial derivative $\partial_x g_-(x,\omega)$, it follows

$$\begin{cases} \begin{pmatrix} A_1 \\ B_1 \end{pmatrix} = \frac{1}{2} \begin{pmatrix} 1 & \frac{1}{n_h} \\ 1 & -\frac{1}{n_h} \end{pmatrix} \begin{pmatrix} 1 & 1 \\ n_0 & -n_0 \end{pmatrix} \begin{pmatrix} A_0 \\ B_0 \end{pmatrix} \\ \begin{pmatrix} A_{2k} \\ B_{2k} \end{pmatrix} = \frac{1}{2} \begin{pmatrix} e^{-in_l \frac{\omega}{c}[(k-1)(h+l)+h]} & \frac{1}{n_l} e^{-in_l \frac{\omega}{c}[(k-1)(h+l)+h]} \\ e^{in_l \frac{\omega}{c}[(k-1)(h+l)+h]} & -\frac{1}{n_l} e^{in_l \frac{\omega}{c}[(k-1)(h+l)+h]} \end{pmatrix} \begin{pmatrix} e^{in_h \frac{\omega}{c}[(k-1)(h+l)+h]} & e^{-in_h \frac{\omega}{c}[(k-1)(h+l)+h]} \\ n_h e^{in_h \frac{\omega}{c}[(k-1)(h+l)+h]} & -n_h e^{-in_h \frac{\omega}{c}[(k-1)(h+l)+h]} \end{pmatrix} \begin{pmatrix} A_{2k-1} \\ B_{2k-1} \end{pmatrix} \\ \begin{pmatrix} A_{2k+1} \\ B_{2k+1} \end{pmatrix} = \frac{1}{2} \begin{pmatrix} e^{-in_h \frac{\omega}{c}k(h+l)} & \frac{1}{n_h} e^{-in_h \frac{\omega}{c}k(h+l)} \\ e^{in_h \frac{\omega}{c}k(h+l)} & -\frac{1}{n_h} e^{in_h \frac{\omega}{c}k(h+l)} \end{pmatrix} \begin{pmatrix} e^{in_l \frac{\omega}{c}k(h+l)} & e^{-in_l \frac{\omega}{c}k(h+l)} \\ n_l e^{in_l \frac{\omega}{c}k(h+l)} & -n_l e^{-in_l \frac{\omega}{c}k(h+l)} \end{pmatrix} \begin{pmatrix} A_{2k} \\ B_{2k} \end{pmatrix} \\ \begin{pmatrix} A_{2N+2} \\ B_{2N+2} \end{pmatrix} = \frac{1}{2} \begin{pmatrix} e^{-in_0 \frac{\omega}{c}L} & \frac{1}{n_0} e^{-in_0 \frac{\omega}{c}L} \\ e^{in_0 \frac{\omega}{c}L} & -\frac{1}{n_0} e^{in_0 \frac{\omega}{c}L} \end{pmatrix} \begin{pmatrix} e^{in_h \frac{\omega}{c}L} & e^{-in_h \frac{\omega}{c}L} \\ n_h e^{in_h \frac{\omega}{c}L} & -n_h e^{-in_h \frac{\omega}{c}L} \end{pmatrix} \begin{pmatrix} A_{2N+1} \\ B_{2N+1} \end{pmatrix} \end{cases} \quad (A.3)$$

while the continuity conditions for $g_+(x,\omega)$ are similar to (A.3), but the coefficients $C_k(\omega)$ and $D_k(\omega)$ replace $A_k(\omega)$ and $B_k(\omega)$ with $k = 0,1,\ldots,2N+1,2N+2$. The $B_0(\omega)$ coefficient is fixed choosing a normalization condition and all the $[A_k(\omega), B_k(\omega)]$ couples with $k = 1,\ldots,2N+2$ are determined applying the continuity conditions (A.3). Similarly for the $C_{2N+2}(\omega)$ coefficient and the $[C_k(\omega), D_k(\omega)]$ couples with $k = 0,1,\ldots,2N+1$.

The QNM frequencies can be calculated, if suitable conditions are imposed to the $g_\pm(x,\omega)$ coefficients. At QNM frequencies $\omega = \omega_n$, where $\text{Im}\,\omega_n < 0$, the auxiliary functions $g_\pm(x,\omega)$ are linearly dependent $g_-(x,\omega) \propto g_+(x,\omega)$; they are "right to left waves" for $x < 0$, so $C_0(\omega_n) = 0$, and "left to right waves" for $x > L$, so:

$$\begin{cases} B_{2N+2}(\omega_n) = 0 \\ A_{2N+2}(\omega_n) = C_{2N+2}(\omega_n) \end{cases}. \quad (A.4)$$

Finally, the QNM norm can be calculated for the stratified medium with refractive index (A.1). The Wronskian $W(x,\omega)$ of the auxiliary functions $g_\pm(x,\omega)$ is obtained from eq. (2.13), using eq. (A.2)

$$W(x,\omega) = \begin{cases} 2in_0 \dfrac{\omega}{c} B_0(\omega)C_0(\omega) & , \quad x \in L_0 \\ 2in_h \dfrac{\omega}{c}[-A_{2k-1}(\omega)D_{2k-1}(\omega) + B_{2k-1}(\omega)C_{2k-1}(\omega)], & x \in L_{2k-1}, \; k=1,\ldots,N+1 \\ 2in_l \dfrac{\omega}{c}[-A_{2k}(\omega)D_{2k}(\omega) + B_{2k}(\omega)C_{2k}(\omega)] & , \quad x \in L_{2k}, \; k=1,\ldots,N \\ 2in_0 \dfrac{\omega}{c} B_{2N+2}(\omega)C_{2N+2}(\omega) & , \quad x \in L_{2N+2} \end{cases} \quad (A.5)$$

The Wronskian $W(x,\omega)$ is x-independent, for the equivalence of all the expressions (A.5), as it can be proved using eq. (A.3), i.e.

$$W(\omega) = 2in_0 \frac{\omega}{c} B_{2N+2}(\omega)C_{2N+2}(\omega), \quad \forall x \in \mathbb{R}. \quad (A.6)$$

The QNM norm $\langle f_n | f_n \rangle = (dW/d\omega)_{\omega=\omega_n}$ is obtained from eq. (A.6), using eq. (A.4), so:

$$\langle f_n | f_n \rangle = 2in_0 \frac{\omega_n}{c} A_{2N+2}(\omega_n)\left(\frac{dB_{2N+2}}{d\omega}\right)_{\omega=\omega_n}. \quad (A.7)$$

## Acknowledgements

A. Settimi is particularly indebted to C. Sibilia and M. Bertolotti who proposed him the strand of research about Quasi Normal Modes. Furthermore, the authors are sincerely grateful to A. Napoli and A. Messina for the interesting discussions on QNMs .

## References.

[19] Since the eigenfunction expansion is intimately related to the dynamics, we consider the causal Green function, defined by [13]

$$[\rho(x)\partial_t^2 - \partial_x^2]G(x,x';t) = \delta(t)\delta(x-x'),\tag{A}$$

together with the initial conditions [13]

$$G(x,x';t) = 0 \text{ for } t \leq 0,\tag{B}$$

$$\rho(x)\partial_t G(x,x';t)\big|_{t=0} = \delta(x-x').\tag{C}$$

Our starting point is the result [5] that under the discontinuity and no tail conditions stated in section 2, the Green function [eq. (A)] can be represented in terms of the eigenfunctions $f_n^N(x)$, defined in section 3, as:

$$G(x,x';t) = \frac{i}{2}\sum_n \frac{f_n^N(x)f_n^N(x')}{\omega_n}e^{-i\omega_n t} \quad \text{for } x,x' \in C, \ t \geq 0.\tag{D}$$

Specifically, this holds even for $x,x' = L$, provided that $t \geq 0$.

Given this representation of the Green function [eq. (D)] and the initial conditions [eq. (B) and eq. (D)], we immediately obtain:

$$\sum_n \frac{f_n^N(x)f_n^N(x')}{\omega_n} = 0,\tag{E}$$

$$\frac{\rho(x)}{2}\sum_n f_n^N(x)f_n^N(x') = \delta(x-x'),\tag{F}$$

Both eq. (E) and eq. (F) hold only for $x,x' \in C$.

**Figures and captions**

Figure 1.a.

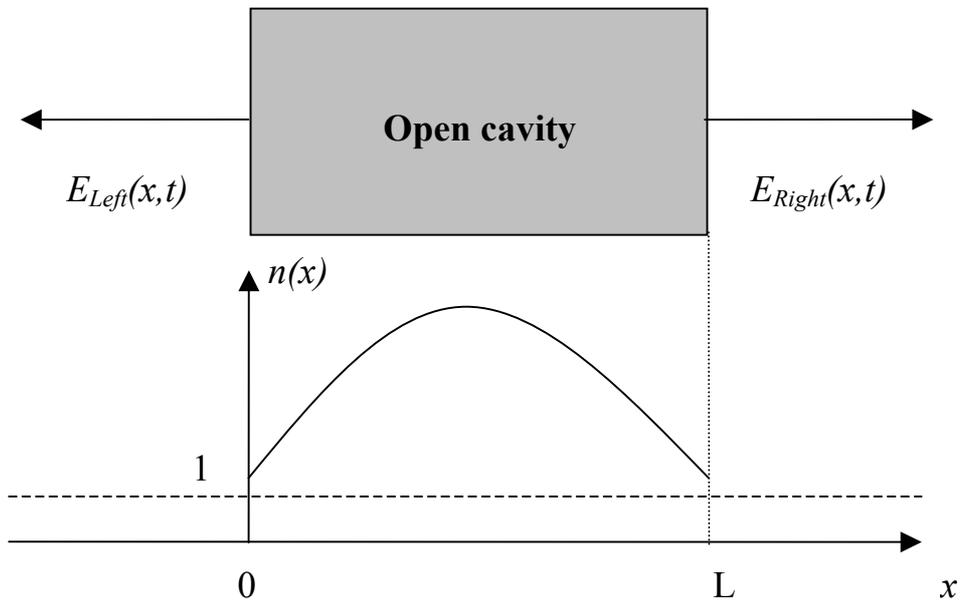

Figure 1.b.

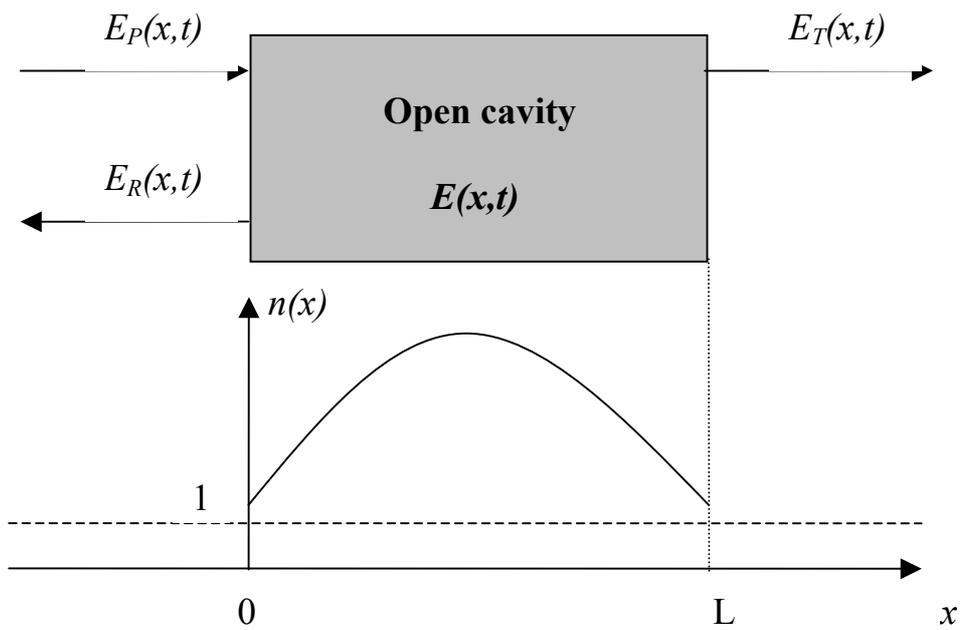

Figure 2.

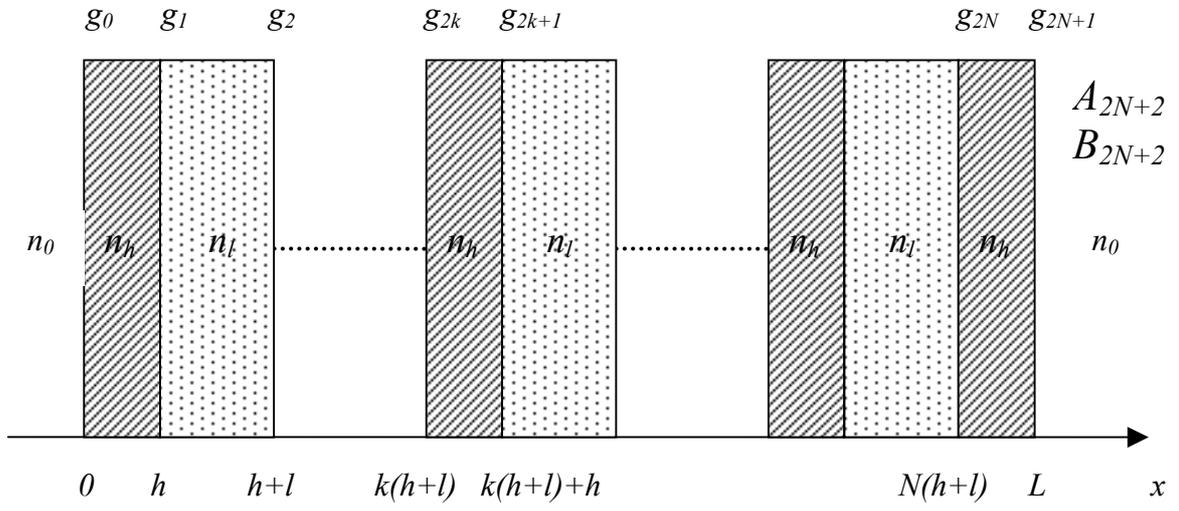

Figure 3.a.

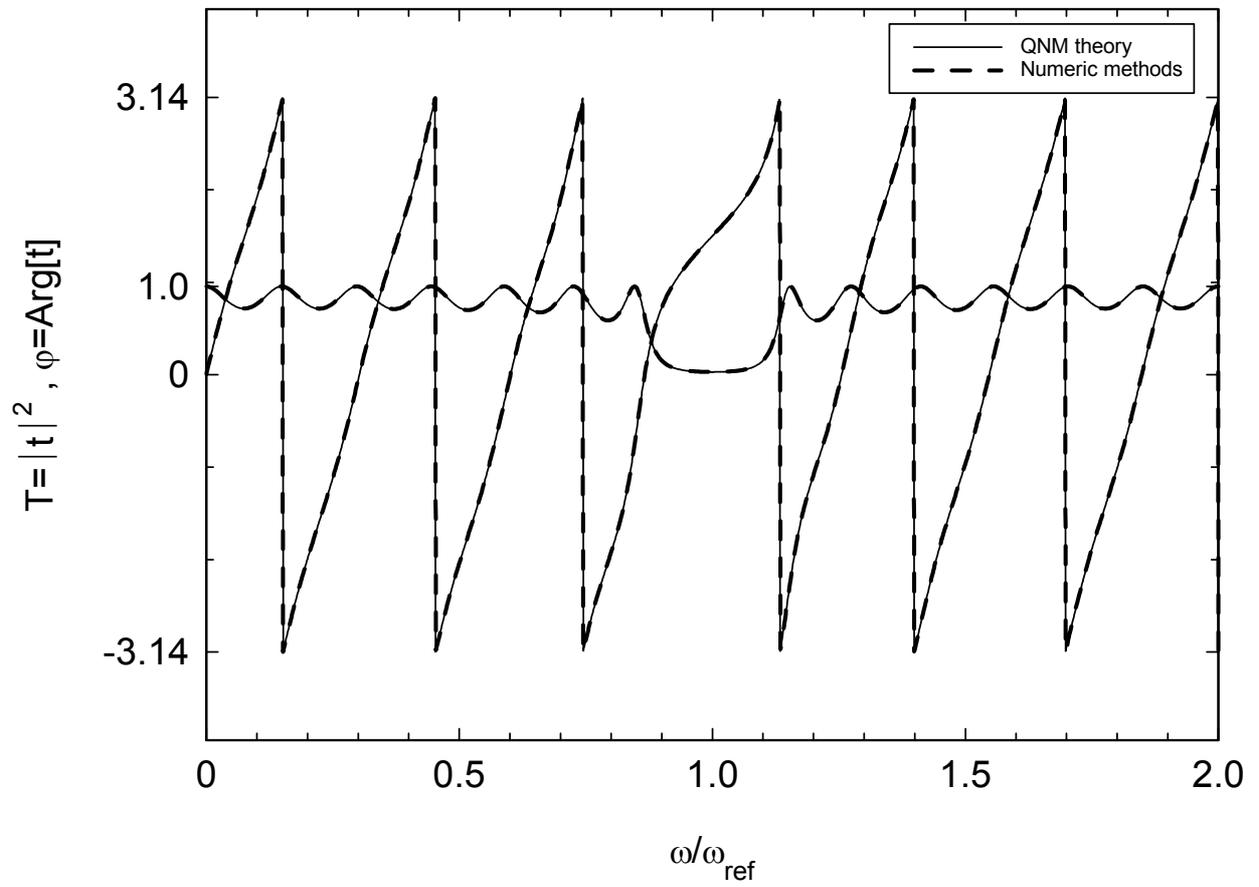

Figure 3.b.

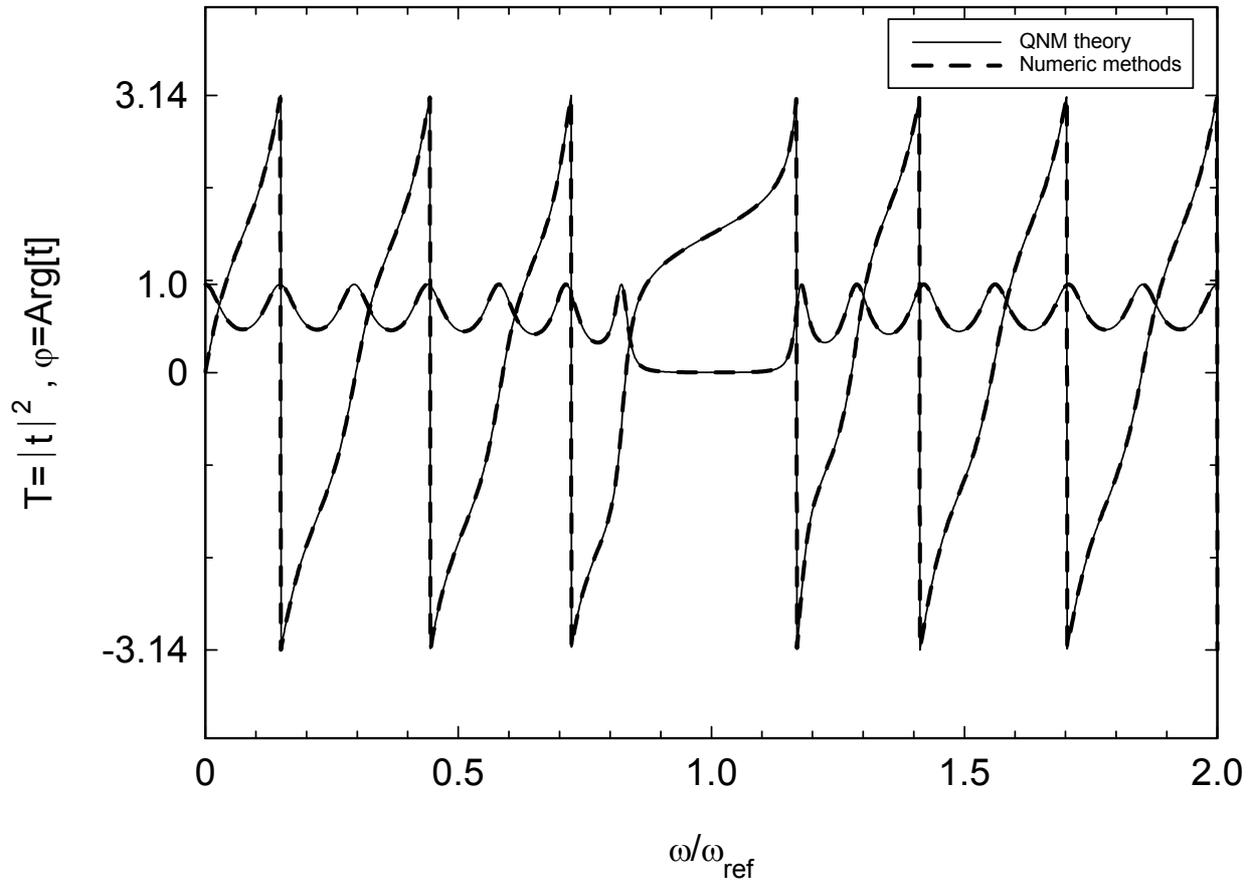

Figure 3.c.

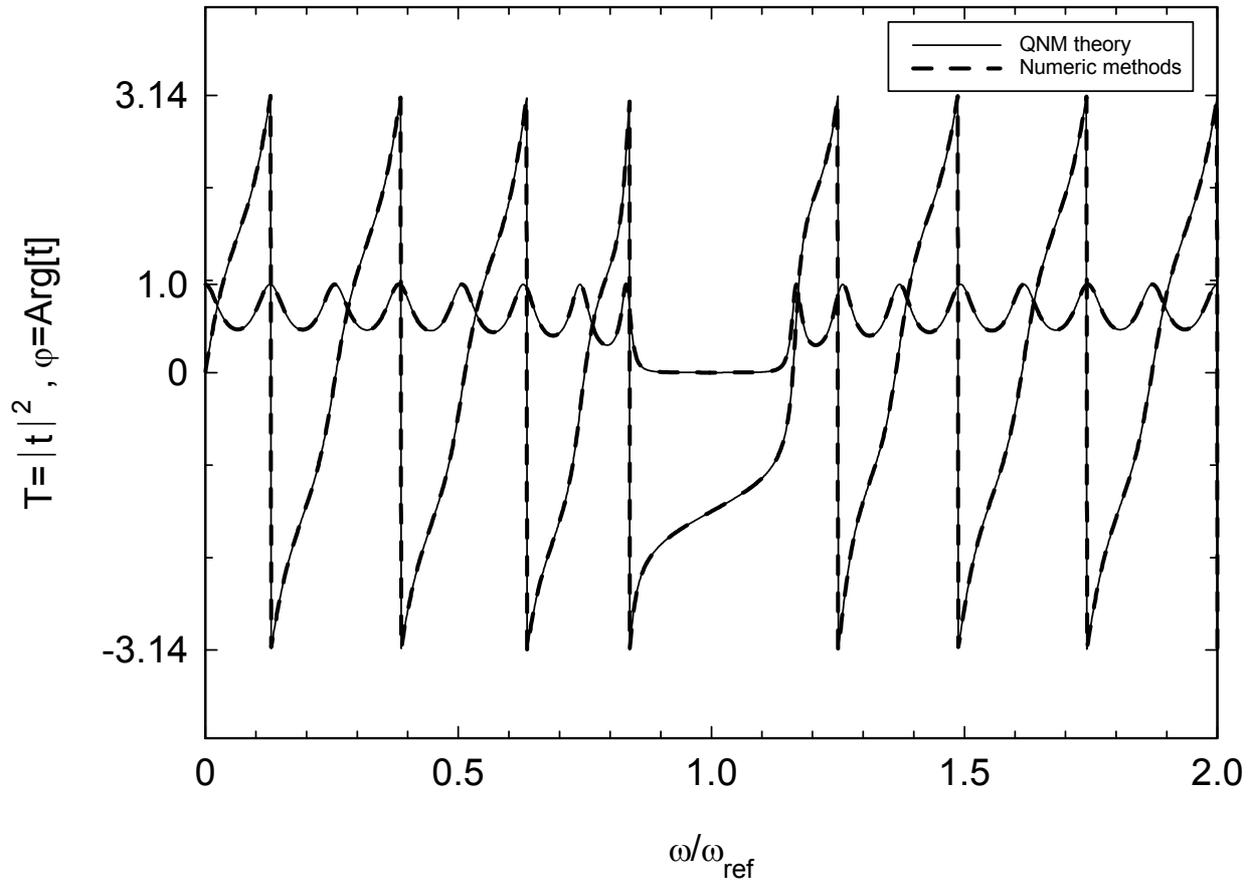

Table 1.a.

| $\lambda_{ref}=1\mu m$, N=6, $n_h=2$, $n_l=1.5$ | Re($\omega_{QNM}/\omega_{ref}$) | $\omega_{Band\text{-}Edge}/\omega_{ref}$ | (Re$\omega_{QNM}$-$\omega_{B.E.}$)/$\omega_{B.E.}$ | $\angle t(\omega_{B.E.}/\omega_{ref})$ |
|---|---|---|---|---|
| Low frequency band-edge | 0.834 | 0.846 | -0.014 | -0.8011 |
| High frequency band-edge | 1.166 | 1.154 | 0.010 | -2.341 |

Table 1.b.

| $\lambda_{ref}=1\mu m$, N=6, $n_h=3$, $n_l=2$ | Re($\omega_{QNM}/\omega_{ref}$) | $\omega_{Band\text{-}Edge}/\omega_{ref}$ | (Re$\omega_{QNM}$-$\omega_{B.E.}$)/$\omega_{B.E.}$ | $\angle t(\omega_{B.E.}/\omega_{ref})$ |
|---|---|---|---|---|
| Low frequency band-edge | 0.825 | 0.822 | 0.003 | -0.6342 |
| High frequency band-edge | 1.175 | 1.178 | -0.002 | -2.507 |

Table 1.c.

| $\lambda_{ref}=1\mu m$, N=7, $n_h=3$, $n_l=2$ | Re($\omega_{QNM}/\omega_{ref}$) | $\omega_{Band\text{-}Edge}/\omega_{ref}$ | (Re$\omega_{QNM}$-$\omega_{B.E.}$)/$\omega_{B.E.}$ | $\angle t(\omega_{B.E.}/\omega_{ref})$ |
|---|---|---|---|---|
| Low frequency band-edge | 0.853 | 0.832 | 0.025 | 2.528 |
| High frequency band-edge | 1.147 | 1.168 | -0.015 | 0.6135 |

Figure 4.a.

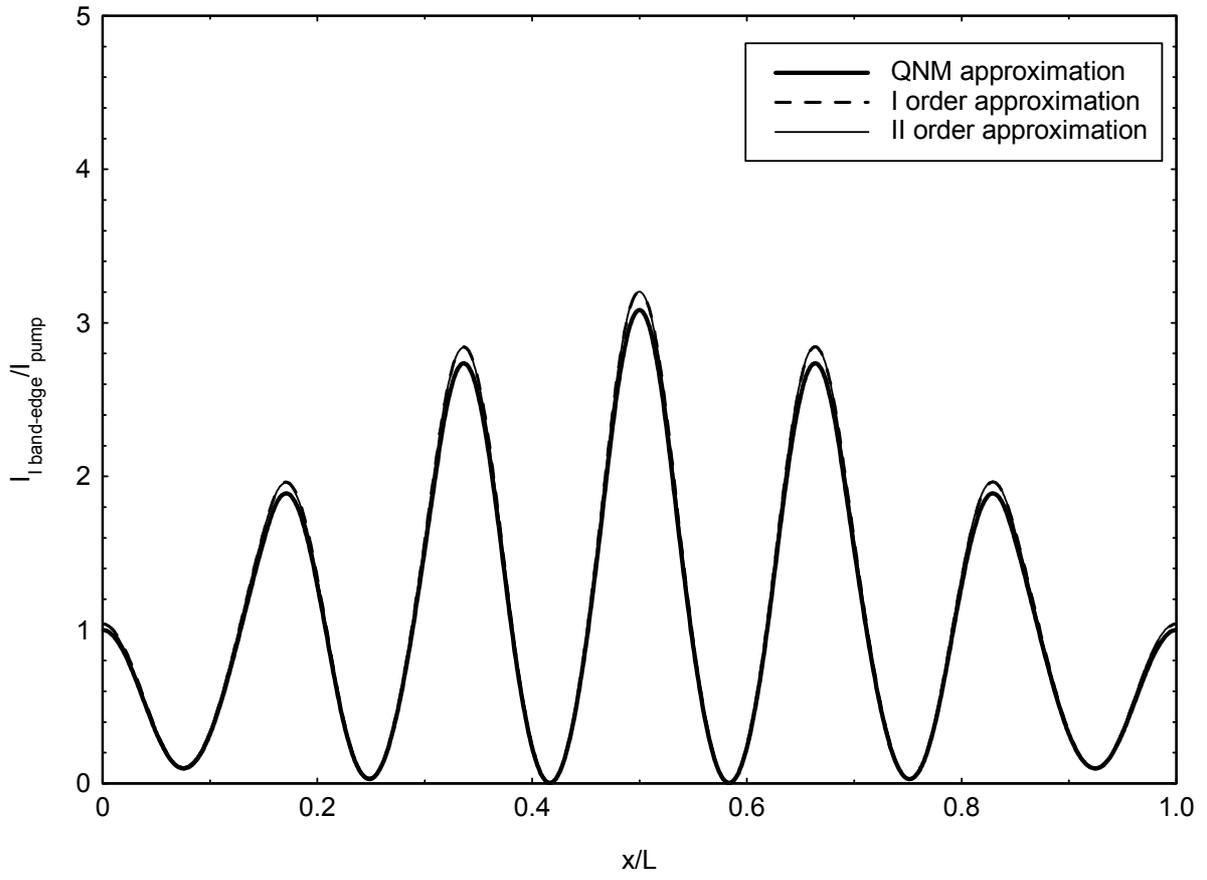

Figure 4.b.

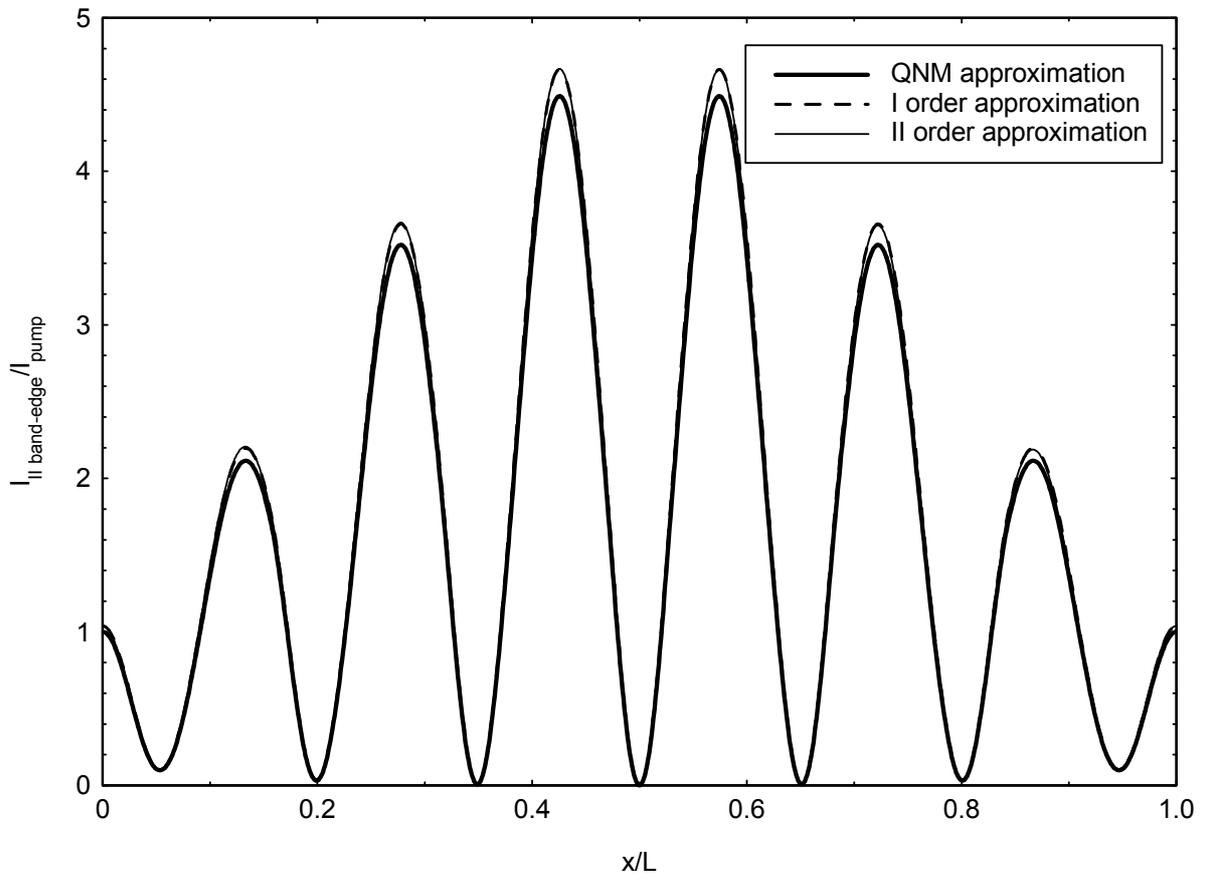

Figure 5.a.

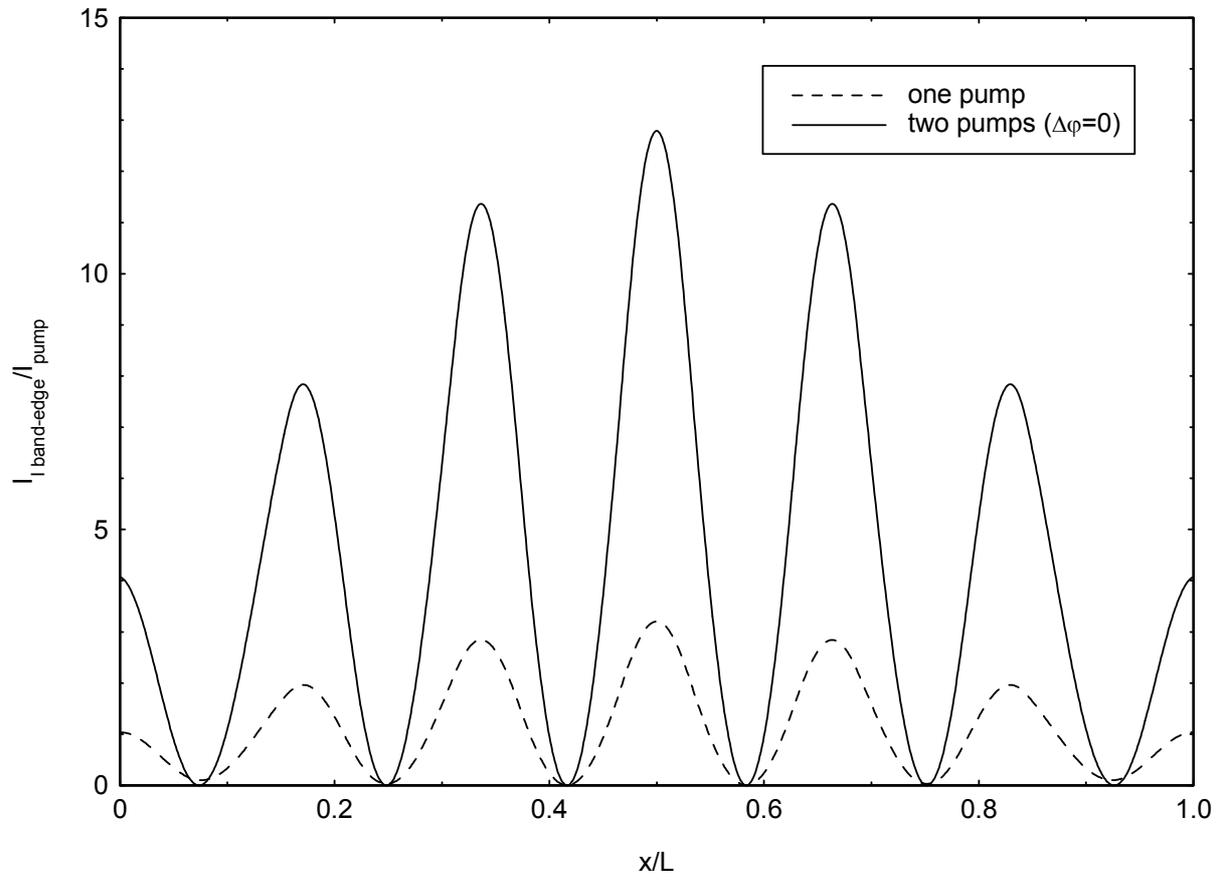

Figure 5.b.

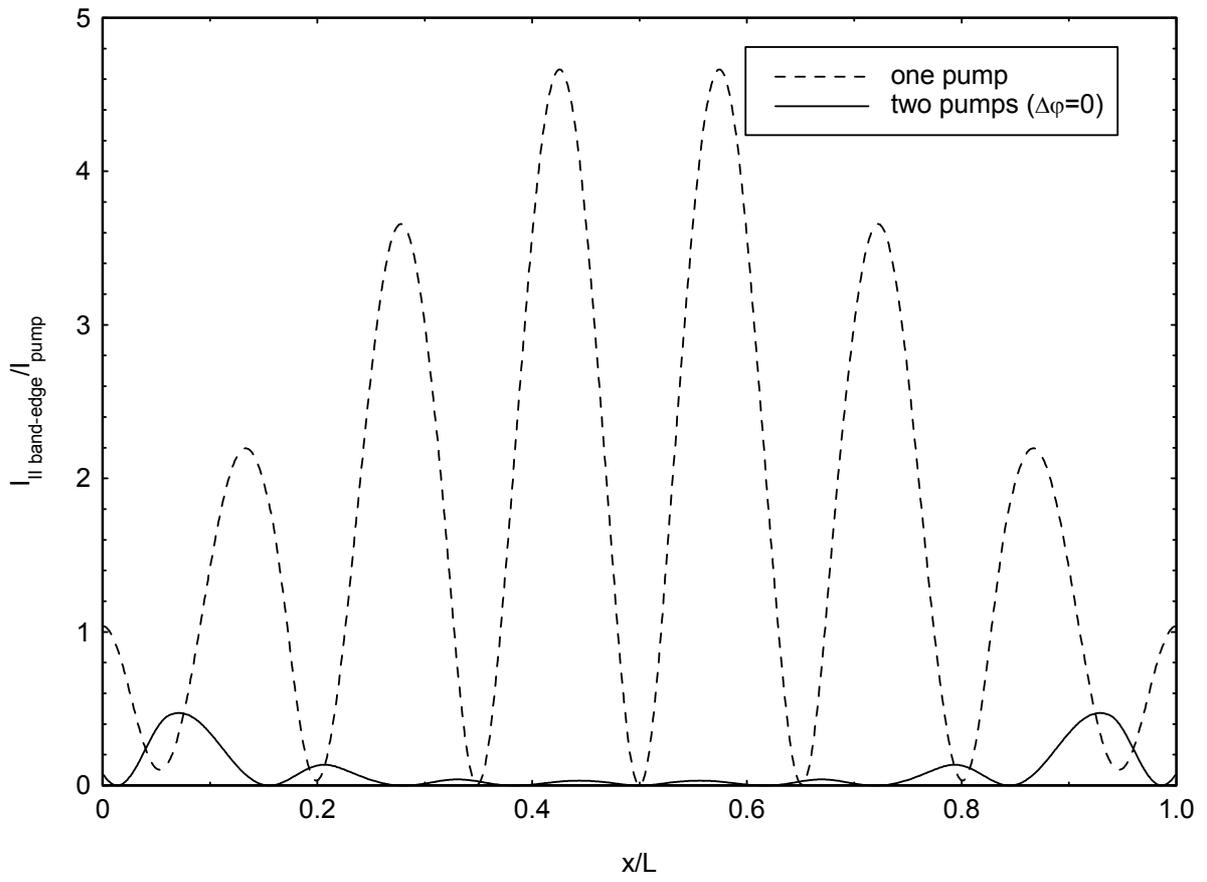

Figure 1. A cavity opened at both ends, of length $L$, with a refractive index $n(x)$: (a) in absence of external pumping, so electromagnetic field satisfies the "outgoing waves" conditions (2.2); (b) in presence of a left-pump $E_P(x,t)$ which satisfies the "incoming wave condition" (4.3).

Figure 2. Refractive index $n(x)$ for a symmetric 1D-PBG structure with $N$ periods plus one layer; every period is composed of two layers, respectively with lengths $h$ and $l$ and with refractive indices $n_h$ and $n_l$, while the added layer is with parameters $h$ and $n_h$.

Figure 3. Transmission spectrum (squared modulus and phase) predicted by QNM theory (—) and by numeric methods existing in the literature [16] (- - -), for QW symmetric 1D-PBG structures, with reference wavelength $\lambda_{ref} = 1\mu m$ and: (a) number of periods $N = 6$, refractive indices $n_h = 2$, $n_l = 1.5$; (b) $N = 6$, refractive indices $n_h = 3$, $n_l = 2$; (c) $N = 7$, $n_h = 3$, $n_l = 2$.

Table 1. Three examples of QW symmetric 1D-PBG structures are considered: (a) $\lambda_{ref} = 1\mu m$, $N = 6$, $n_h = 2$, $n_l = 1,5$, (b) $\lambda_{ref} = 1\mu m$, $N = 6$, $n_h = 3$, $n_l = 2$, (c) $\lambda_{ref} = 1\mu m$, $N = 7$, $n_h = 3$, $n_l = 2$. For each of the three examples, the low and high frequency band-edges are described in terms of their resonances ($\omega_{Band-Edge}/\omega_{ref}$) and phases $\angle t(\omega_{B.E.}/\omega_{ref})$; besides, there is one QNM close to every single band-edge: the real part of the QNM-frequency $\mathrm{Re}(\omega_{QNM}/\omega_{ref})$ is reported together with the relative shift from the band-edge resonance $(\mathrm{Re}\,\omega_{QNM} - \omega_{B.E.})/\omega_{B.E.}$.

Figure 4. With reference to a QW symmetric 1D-PBG structure ($\lambda_{ref} = 1\mu m$, $N = 6$, $n_h = 3$, $n_l = 2$), the e.m. "mode" intensities (a) $I_{I\,band-edge}$ at the low frequency band edge ($\omega_{I\,band-edge}/\omega_{ref} = 0.822$) and (b) $I_{II\,band-edge}$ at the high frequency band edge ($\omega_{II\,band-edge}/\omega_{ref} = 1.178$), in units of the intensity for an incoming pump $I_{pump}$, are plotted as functions of the dimensionless space $x/L$, where $L$ is the length of the 1D-PBG structure. It is clear the shifting between the QNM approximation (6.10) (the e.m. mode intensity $I_{I\,band-edge}$ is due to QNM close to the band-edge) and the first order approximation (6.6)-(6.7) [more refined than (6.10), $I_{band-edge}$ is calculated as the product of the QNM, close to the frequency band-edge, for a weigh coefficient, which takes into account the shift between the band-edge resonance and the QNM frequency] or the second order approximation (6.8)-(6.9) [even more refined, but almost

superimposed to (6.6)-(6.7), $I_{band-edge}$ is calculated as the sum of the first order contribution due to the QNM, close to the frequency band-edge, and the second order contributions of the two adjacent QNMs, with the same imaginary part, so belonging to the same QNM family].

Figure 5. With reference to a QW symmetric 1D-PBG structure ($\lambda_{ref} = 1\mu m$, $N = 6$, $n_h = 3$, $n_l = 2$), the e.m. "mode" intensities excited inside the open cavity by one pump (- - -) coming from the left side [I order approximation (6.6)] and by two counter-propagating pumps in phase (—) [see eq. (7.9)] are compared when each of the two pumps are tuned at (a) the low frequency band-edge ($\omega_{I\ band-edge}/\omega_{ref} = 0.822$) or (b) the high frequency band-edge ($\omega_{II\ band-edge}/\omega_{ref} = 1.178$). The e.m. mode intensities $I_{I\ band-edge}$ and $I_{II\ band-edge}$, in units of the intensity for the two pumps $I_{pump}$ are plotted as functions of the dimensionless space $x/L$, where $L$ is the length of the 1D-PBG structure. Thus, two counter-propagating pumps tuned at the same transmission resonance do not necessarily excite the cavity mode which one might expect at that frequency if the same pumps have an appropriate phase-difference.